\documentclass[reprint, superscriptaddress, amsmath, amssymb, aps, pra, longbibliography, floatfix]{revtex4-2}

\usepackage{amsmath}
\usepackage{graphicx}
\usepackage{braket}
\usepackage{amssymb}
\usepackage{bm}
\usepackage{silence}
\usepackage[colorlinks=true,linkcolor=blue,anchorcolor=blue,citecolor=blue,urlcolor=blue]{hyperref}

\begin{document}
\preprint{APS/123-QED}

\title{\texorpdfstring{Orbit-resolved spin holography: role of Coulomb focusing in target-dependent polarization}{Orbit-resolved spin holography: role of Coulomb focusing in target-dependent polarization}}

\author{Tao Chen}
\affiliation{State Key Laboratory of Dark Matter Physics, Key Laboratory for Laser Plasmas (Ministry of Education) and School of Physics and Astronomy, Collaborative Innovation Center for IFSA (CICIFSA), Shanghai Jiao Tong University, Shanghai 200240, China}
\affiliation{Tsung-Dao Lee Institute, Shanghai Jiao Tong University, Shanghai 201210, China}
\author{Yang Li}
\affiliation{State Key Laboratory of Dark Matter Physics, Key Laboratory for Laser Plasmas (Ministry of Education) and School of Physics and Astronomy, Collaborative Innovation Center for IFSA (CICIFSA), Shanghai Jiao Tong University, Shanghai 200240, China}
\author{Fang Liu}
\affiliation{State Key Laboratory of Dark Matter Physics, Key Laboratory for Laser Plasmas (Ministry of Education) and School of Physics and Astronomy, Collaborative Innovation Center for IFSA (CICIFSA), Shanghai Jiao Tong University, Shanghai 200240, China}
\author{Pei-Lun He} \email{peilunhe@sjtu.edu.cn}
\affiliation{State Key Laboratory of Dark Matter Physics, Key Laboratory for Laser Plasmas (Ministry of Education) and School of Physics and Astronomy, Collaborative Innovation Center for IFSA (CICIFSA), Shanghai Jiao Tong University, Shanghai 200240, China}
\author{Carla Figueira de Morisson Faria} \email{c.faria@ucl.ac.uk}
\affiliation{Department of Physics and Astronomy, University College London, Gower Street, London WC1E 6BT, United Kingdom}
\author{Feng He} \email{fhe@sjtu.edu.cn}
\affiliation{State Key Laboratory of Dark Matter Physics, Key Laboratory for Laser Plasmas (Ministry of Education) and School of Physics and Astronomy, Collaborative Innovation Center for IFSA (CICIFSA), Shanghai Jiao Tong University, Shanghai 200240, China}
\affiliation{Tsung-Dao Lee Institute, Shanghai Jiao Tong University, Shanghai 201210, China}

\date{\today}

\begin{abstract}
Strong-field photoelectron holography encodes ultrafast electron dynamics through momentum-space interference. However, the orbit-resolved origin of spider-like spin fringes and the mechanism underlying their target dependence remain unclear. Here, we resolve both issues by analyzing photoelectron spin textures generated during tunneling ionization. We use the Coulomb quantum-orbit strong-field approximation, benchmarked against time-dependent Schr\"odinger equation simulations for $\mathrm{He^+}$ and Xe, to separate orbital-channel and quantum-orbit contributions. Spider-like fringes arise from interference between $p$-orbital ionization channels with different magnetic quantum numbers within an individual orbit class and therefore do not require interorbit interference. The observable polarization along these fringes, however, depends on the balance among orbit-class contributions. The decomposition associates the opposite first-leg polarizations of $\mathrm{He^+}$ and Xe with different relative weights of laser-deflected and forward-scattered trajectories, consistent with target-dependent Coulomb focusing. Photoelectron spin textures thus complement momentum distributions as probes of Coulomb-driven strong-field dynamics.
\end{abstract}

\maketitle

\section{Introduction}
The development of intense ultrashort laser sources has established strong-field laser-matter interactions as a powerful means of probing electronic dynamics and structure in atoms and molecules on intrinsic subfemtosecond timescales~\cite{RevModPhys.81.163, Salieres_2012}. In the low-frequency, high-intensity regime, ionization can proceed by tunneling through the field-suppressed Coulomb barrier. After tunneling, the liberated electron is driven by the oscillating laser field and may revisit the parent ion, producing a range of recollision processes~\cite{PhysRevLett.71.1994}. These include radiative recombination leading to high-order harmonic generation~\cite{PhysRevA.49.2117}, elastic rescattering responsible for high-order above-threshold ionization~\cite{PhysRevA.51.1495, G_G_Paulus_1994}, and inelastic recollision that can trigger multielectron dynamics such as nonsequential double ionization~\cite{PhysRevLett.71.1994, RevModPhys.84.1011}. These recollision processes support ultrafast imaging through high-harmonic spectroscopy~\cite{PhysRevLett.94.053004, PhysRevLett.133.023201}, laser-induced electron diffraction~\cite{doi:10.1126/science.1157980, Blaga2012, PhysRevX.14.011015, Pullen2015}, and strong-field photoelectron holography (SFPH)~\cite{doi:10.1126/science.1198450, PhysRevLett.109.013002}. In SFPH, interference between electron wave packets released at different times or following different continuum paths encodes information about the ionic potential and subcycle electron motion.

SFPH is a laser-driven analog of optical holography in which direct and rescattered electron wave packets provide reference and signal pathways. Several characteristic patterns have been identified~\cite{PhysRevA.84.043420, Li:16, PhysRevA.102.033111}. The \emph{spider} arises from interference between laser-deflected and forward-scattered trajectories released within the same quarter cycle, whereas the \emph{fishbone} involves laser-deflected and backward-scattered trajectories released in different quarter cycles~\cite{Khurelbaatar2024}. These patterns provide time-resolved access to electron emission dynamics and, in some cases, to atomic or molecular structure~\cite{PhysRevA.104.013109, PhysRevLett.116.133001, PhysRevA.102.013109, wcl3-x52t, PhysRevLett.108.263003}. The particularly robust spider pattern has been observed in numerous experiments~\cite{doi:10.1126/science.1198450, PhysRevLett.109.013002, PhysRevLett.109.073004, Figueira}. Molecular orientation and nodal planes can modulate the observed spider pattern~\cite{Meckel2014, PhysRevLett.120.133204, Walt2017}. However, because its two principal pathways are released within the same half cycle, their phase difference produces broad fringes with limited intrinsic sensitivity to target parity~\cite{PhysRevA.102.013109}. Parity sensitivity can emerge when a third pathway from another half cycle modulates the spider~\cite{PhysRevA.104.013109}. The prominent spider can also mask subtler structure-sensitive features, including the fishbone in the same momentum region~\cite{PhysRevLett.108.263003}. These limitations motivate complementary observables that make the dominant spider feature sensitive to target-dependent Coulomb dynamics.

Conventional SFPH extracts holographic information from the photoelectron momentum distribution (PMD). Spin-resolved detection provides access to an internal degree of freedom and can expose correlations between orbital ionization channels that are not explicit in the spin-integrated distribution. At nonrelativistic intensities, spin-orbit coupling (SOC) is weak during continuum propagation, but bound-state SOC can still correlate the orbital and spin degrees of freedom. Early studies therefore examined how this intrinsic atomic coupling shapes spin-resolved single- and multiphoton ionization~\cite{PhysRev.178.131, PhysRevLett.30.413, PhysRevA.88.013401, Hartung2016, PhysRevLett.120.043202}, including spin polarization in circularly polarized fields~\cite{PhysRevA.95.063410, PhysRevA.88.013401} and spin-dependent above-threshold ionization spectra~\cite{PhysRevLett.120.043201}. Recent work demonstrated vortex-like photoelectron spin textures (PSTs) in linearly polarized tunneling ionization and introduced spin-polarized photoelectron holography based on direct-rescattered interference~\cite{PhysRevLett.134.163201}. A subsequent study showed that relative phases among $p$-orbital ionization channels generate a toroidal PST in circularly polarized fields~\cite{mao2026ultrafast}. These results leave unresolved whether a spider-like PST fringe can arise within a single orbit class and what controls its target-dependent polarization sign. An orbit-resolved trajectory analysis is needed to separate the interchannel fringe mechanism from target-dependent orbit-class weighting~\cite{Figueira, PhysRevA.110.033108}.

Here, we combine the Coulomb quantum-orbit strong-field approximation (CQSFA)~\cite{PhysRevA.92.043407, Carlsen_2024} with orbit classification~\cite{PhysRevLett.105.253002} and benchmark the resulting calculations against time-dependent Schr\"odinger equation (TDSE) simulations. Building on the spin holography introduced in Ref.~\cite{PhysRevLett.134.163201}, we provide an orbit-resolved account of its spider-like fringes. Within an individual orbit class, the spider geometry arises from interference between ionization channels with different orbital magnetic quantum numbers and therefore does not require interorbit interference. For the two targets considered, the PMD spiders remain qualitatively similar, whereas the PST polarization along the first spider leg reverses sign. The orbit decomposition associates this reversal with different relative weights of laser-deflected trajectories in orbit class 2 and forward-scattered trajectories in orbit class 3, consistent with target-dependent Coulomb focusing. These findings suggest that PSTs can provide target-sensitive probes of Coulomb-driven strong-field dynamics.

The paper is organized as follows. Sec.~\ref{secii} provides a brief derivation of the PST. In Sec.~\ref{seciii}, we describe the TDSE simulations and CQSFA calculations. Sec.~\ref{seciv} presents the PSTs and PMDs for $\mathrm{He^+}$ and Xe, together with an analysis based on orbit classification. Sec.~\ref{secv} concludes the paper. Atomic units, with $\hbar = e = m_e = 1$, are used throughout unless stated otherwise.

\section{Photoelectron spin texture \label{secii}}
We briefly review the PST introduced in Ref.~\cite{PhysRevLett.134.163201} and establish the notation used below.
Taking the $z$ axis as the quantization axis, we represent the spin-up and spin-down basis states by $\ket{\uparrow}=\left(1\ 0\right)^{\mathrm{T}}$ and $\ket{\downarrow}=\left(0\ 1\right)^{\mathrm{T}}$, respectively. A pure spin state is $\chi_u\ket{\uparrow}+\chi_d\ket{\downarrow}$, where $\chi_u$ and $\chi_d$ are complex amplitudes. When spatial degrees of freedom are included, these amplitudes become spatial wavefunction components, denoted by $\ket{\psi_u^{jm}}$ and $\ket{\psi_d^{jm}}$. We consider ionization from atomic $np$ orbitals.
Within the single-active-electron approximation, SOC splits an $np$ orbital into the $j=1/2$ and $j=3/2$ manifolds. In the $j=1/2$ manifold, each $p_{\pm}$ component is associated with the opposite spin projection. This correlation enables efficient spin polarization in circularly polarized fields through preferential ionization of one $p_{\pm}$ component~\cite{PhysRevA.88.013401}. In the $j=3/2$ manifold considered here, the same $p_{\pm}$ component occurs with both spin projections across the $m$ sublevels. To evaluate the CQSFA ionization dipole analytically, we represent the initial $np_j$ state by a hydrogenlike radial function multiplied by a spinor spherical harmonic,
\begin{equation}
    \mathbf{\Psi}_{\kappa njm}(\bm{r}) = C_{\kappa n}rL_{n-2}^{3}(2\kappa r/n)\text{e}^{-\kappa r/n}\mathbf{\Omega}_{jm}. \label{psi0}
\end{equation}
Here, $C_{\kappa n}$ normalizes the hydrogenlike radial function, $L_{n-2}^3(x)$ is an associated Laguerre polynomial, and $\mathbf{\Omega}_{jm}$ is a spinor spherical harmonic. The symbol $\kappa=\sqrt{2n^2I_p}$ denotes the model effective charge. The corresponding radial exponent is $\kappa/n=\kappa_b=\sqrt{2I_p}$, where $\kappa_b$ is the bound-state momentum used in the tunnel-exit formula below. This model state is used only for the CQSFA ionization dipole; the TDSE orbitals are obtained independently from the target-specific field-free Hamiltonian.
For $j=3/2$, these spinors are given by
\begin{equation}
    \begin{aligned}
        \bm{\Omega}_{\frac{3}{2},\frac{1}{2}} &= 
        \begin{pmatrix}
        \sqrt{\tfrac{2}{3}}\,Y_{1,0} \\
        \sqrt{\tfrac{1}{3}}\,Y_{1,1}
        \end{pmatrix}, 
        \bm{\Omega}_{\frac{3}{2},-\frac{1}{2}} = 
        \begin{pmatrix}
        \sqrt{\tfrac{1}{3}}\,Y_{1,-1} \\
        \sqrt{\tfrac{2}{3}}\,Y_{1,0}
        \end{pmatrix}, \\
        \bm{\Omega}_{\frac{3}{2},\frac{3}{2}} &= 
        \begin{pmatrix} Y_{1,1} \\ 0 \end{pmatrix}, \qquad
        \bm{\Omega}_{\frac{3}{2},-\frac{3}{2}} = 
        \begin{pmatrix} 0 \\ Y_{1,-1} \end{pmatrix},
    \end{aligned} \label{CG}
\end{equation}
The Clebsch--Gordan coefficients therefore connect the orbital subchannels to the spin basis.
For the laser parameters considered, we neglect continuum SOC and relativistic spin precession~\cite{PhysRevLett.2.435, PhysRevA.110.033108}. The spin-up and spin-down components then propagate independently. The transition amplitude for final momentum $\bm{p}$ and initial quantum numbers $(j,m)$ is
\begin{equation}
    \begin{aligned}
            & \bm{M}(\bm{p};j,m) \\
            & =-\text{i}\int_{t_i}^{t_f}dt
                \begin{pmatrix}
                    \langle\Psi_{\bm{p}}|U(t_f,t)H_I(t)U_0(t,t_i)|\psi_u^{jm}\rangle \\
                    \langle\Psi_{\bm{p}}|U(t_f,t)H_I(t)U_0(t,t_i)|\psi_d^{jm}\rangle
                \end{pmatrix},
    \end{aligned} \label{transition_amp}
\end{equation}
where $\ket{\psi_u^{jm}}$ and $\ket{\psi_d^{jm}}$ are the spatial components of the initial spinor. The propagators are $U(t_2,t_1)=\mathcal{T}\exp[-\mathrm{i}\int_{t_1}^{t_2}H(\tau)d\tau]$ and $U_0(t_2,t_1)=\exp[-\mathrm{i}H_0(t_2-t_1)]$. Here, $H(t)=H_0+H_I(t)$, with $H_0=\bm{p}^2/2+V_0(r)$ and $H_I(t)=\bm{r}\cdot\bm{E}(t)$ in the length gauge, and $\bm{E}(t)=-\partial_t\bm{A}(t)$. When the initial $m$ sublevels are equally populated and the residual-ion magnetic substate is unresolved, the observable PST is obtained by an incoherent sum over $m$,
\begin{equation}
\langle\bm{\zeta}(\bm{p};j)\rangle=\frac{\sum_{m}\bm{M}^\dagger(\bm{p};j,m)\bm{\sigma}\bm{M}(\bm{p};j,m)}{\sum_{m}\bm{M}^\dagger(\bm{p};j,m)\bm{M}(\bm{p};j,m)}, \label{spin-polarization}
\end{equation}
where $\bm{\sigma}$ is the vector of Pauli matrices. After the incoherent sum over $m$, the physical PST does not depend on the arbitrary quantization-axis representation, although its Cartesian components transform covariantly under rotations.

For $j=3/2$, the PST can be expressed compactly in terms of the orbital-resolved ionization amplitudes $\chi^{(0)}$, $\chi^{(+)}$, and $\chi^{(-)}$.
These amplitudes are the scalar transition amplitudes obtained from Eq.~\eqref{transition_amp} when the initial spatial component is taken to be the corresponding $p_0$ or $p_\pm$ orbital,
\begin{equation}
    \begin{aligned}
        \chi^{(\mu)}(\bm{p})
        &=-\text{i}\int_{t_i}^{t_f}dt\,
        \langle\Psi_{\bm{p}}|U(t_f,t)H_I(t) \\
        &\quad \times U_0(t,t_i)|p_\mu\rangle,
        \qquad \mu=0,+,-.
    \end{aligned}
\end{equation}
Here, $|p_\mu\rangle$ denotes the radial $np$ wavefunction multiplied by the spherical harmonic $Y_{1,\mu}$.
Adopting the convention
\begin{equation}
    \begin{aligned}
        & \chi^{(x)}=\frac{1}{\sqrt{2}}\big(\chi^{(+)}-\chi^{(-)}\big), \\
        & \chi^{(y)}=\frac{\text{i}}{\sqrt{2}}\big(\chi^{(+)}+\chi^{(-)}\big),
    \end{aligned}
\end{equation}
and defining $N=|\chi^{(+)}|^2+|\chi^{(-)}|^2+|\chi^{(0)}|^2$, we can explicitly express the PST as
\begin{equation}
    \begin{aligned}
            & \langle\bm{\zeta}(\bm{p};j=\frac{3}{2})\rangle \\
            & =\frac{1}{N}\bigg(\text{Im}\big[\chi^{(0)*}\chi^{(y)}\big],\text{Im}\big[\chi^{(0)*}\chi^{(x)}\big],\text{Im}\big[\chi^{(x)*}\chi^{(y)}\big]\bigg). \label{spin}
    \end{aligned}
\end{equation}
Equation~\eqref{spin} shows that each transverse spin component probes coherence between different orbital channels. This interchannel coherence is the basis of the orbit-resolved spin-holographic mechanism analyzed below.

\section{\texorpdfstring{Numerical evaluation of the transition amplitude}{Numerical evaluation of the transition amplitude} \label{seciii}}
We evaluate each orbital amplitude $\chi^{(\mu)}(\bm{p})$ using two complementary approaches: direct TDSE propagation and the CQSFA. The spinor amplitudes and PST are then reconstructed using Eqs.~\eqref{CG} and~\eqref{spin}. Throughout this section, $\ket{\psi_0}$ denotes one selected spatial orbital, $\ket{p_0}$, $\ket{p_+}$, or $\ket{p_-}$, rather than the full spinor state.

All calculations use a linearly polarized, four-cycle laser pulse with a central wavelength of $2000\,\mathrm{nm}$ and vector potential
\begin{equation}
    \bm{A}(t)=A_0\sin(\omega t+\delta)\sin^2\left(\frac{\omega t}{2N_c}\right)\hat{\bm{e}}_x,
\end{equation}
where $A_0$ is the vector-potential amplitude, $\omega$ is the central angular frequency, $\delta$ is the carrier-envelope phase (CEP), and $N_c=4$. The peak intensity is $1.825\times10^{13}\,\mathrm{W/cm^2}$, for which $\gamma\lesssim1$ for both targets. Although the $p_0$ channel contributes weakly to the total yield, it enters transverse PST components through interchannel terms and is retained in the full spin reconstruction.

\subsection{TDSE simulation}
We solve the three-dimensional TDSE in spherical coordinates using the open-source QPC-TDSE code~\cite{ZHANG2023108787}. The wavefunction is expanded in ninth-order B-splines and spherical harmonics,
\begin{equation}
    \psi(\bm{r},t)=\sum_{nlm}c_{nlm}(t)\frac{B_n(r)}{r}Y_{l,m}(\theta,\phi).
\end{equation}
The initial orbital is obtained by diagonalizing the target-specific field-free Hamiltonian. We propagate it with a Crank--Nicolson algorithm combined with a split-operator scheme. After the pulse, the final wavefunction is projected onto exact scattering states with incoming-wave boundary conditions to obtain the continuum amplitude,
\begin{equation}
    M(\bm{p})=\langle\psi_{\bm{p}}^{(-)}|\psi(t_f)\rangle.
\end{equation}
The channel-resolved PMD is $|M(\bm{p})|^2$, and unresolved initial sublevels are combined incoherently.

For $\mathrm{He^+}$, we use the bare Coulomb potential $V_0(r)=-2/r$. For Xe, we use the Tong--Lin effective potential~\cite{Tong_2005},
\begin{equation}
    V_0(r) = -\frac{Z+a_1\text{e}^{-a_2r}+a_3r\text{e}^{-a_4r}+a_5\text{e}^{-a_6r}}{r},
\end{equation}
with the standard parameter set of Ref.~\cite{PhysRevA.90.043410}, which gives $I_p\approx0.4458\,$a.u. for the Xe $5p_{3/2}$ orbital.

The radial grid extends to $r_{\max}=2200$ a.u., so the ionized wave packet remains inside the box through the end of the pulse. We use $N_r=5000$ radial B-splines and $N_l=112$ angular-momentum channels. These basis sizes were checked for numerical convergence. For Xe, a quadratic knot sequence resolves the rapid near-core variation of the effective potential. The Crank--Nicolson time step is $10^{-3}$ a.u. for Xe and $5\times10^{-3}$ a.u. for $\mathrm{He^+}$.

We use excited-state $\mathrm{He^+}$ as a qualitative reference because its bare Coulomb potential isolates residual-charge effects and its $2p$ ionization potential is close to that of Xe. The $2p_{3/2}$--$2p_{1/2}$ fine-structure splitting is approximately $7.3\times10^{-4}\,$eV and is negligible on the energy scales considered here. The $2p$ state could be prepared by resonant extreme-ultraviolet excitation of the $1s$--$2p$ transition near $30.4\,\mathrm{nm}$. High-harmonic excitation of singly ionized helium has been demonstrated in this spectral range~\cite{Gruendeman2024}.

\subsection{CQSFA}
To identify the quantum-orbit origin of the spin-holographic fringes, we also evaluate the transition amplitude with the CQSFA~\cite{PhysRevA.92.043407, Carlsen_2024}.
In Eq.~\eqref{transition_amp}, ionization occurs through $H_I(t)$ at time $t$, followed by continuum propagation in the combined laser and ionic fields. At asymptotically large distances, we approximate the scattering state by the plane wave $\langle\bm{r}|\bm{p}\rangle=(2\pi)^{-3/2}\mathrm{e}^{\mathrm{i}\bm{p}\cdot\bm{r}}$. The scalar orbital amplitude is then
\begin{equation}
    M(\bm{p})= -{\rm{i}} \lim_{t_f\to\infty} \int_{t_i}^{t_f}dt\langle \bm{p}|U(t_f,t)H_I(t)|\psi_0(t)\rangle.
\end{equation}
Here, $\ket{\psi_0(t)}=\exp(\mathrm{i}I_p(t-t_i))\ket{\psi_0}$. The composition law for $U$ gives the phase-space path-integral representation~\cite{kleinert2006path}
\begin{equation}
\begin{split}
    M(\bm{p}) = &-{\rm{i}} \lim_{t_f\to\infty}\int_{t_i}^{t_f}dt \int^{\bm{p}} D\bm{p}' \int_{\bm{r}_0} \frac{D\bm{r}'}{(2\pi)^3} \\
    &\times~e^{{\rm{i}}S(t)} \langle\bm{p}_0(t)|H_I(t)|\psi_0\rangle ,\label{path_integral}
\end{split}
\end{equation}
with the action
\begin{equation}
    S(t) = I_p(t-t_i)-\int_{t}^{t_f}\big(\bm{r}'(\tau)\cdot\dot{\bm{p}'}(\tau)+H(\bm{r}'(\tau),\bm{p}'(\tau),\tau)\big)d\tau,
\end{equation}
where $\bm r'(\tau)$ and $\bm p'(\tau)$ are intermediate phase-space paths. Appendix~\ref{appendix a} evaluates the ionization dipole $d[\bm{p}_0(t),t]=\langle\bm{p}_0(t)|H_I(t)|\psi_0\rangle$ for the model state in Eq.~\eqref{psi0}.

In the semiclassical approximation, the dominant contributions arise from stationary classical paths satisfying
\begin{equation}
    \dot{\bm{r}}_s(\tau)=\bm{p}_s(\tau),
\end{equation}
\begin{equation}
    \dot{\bm{p}}_s(\tau)=-\bm{E}(\tau)-\nabla V_0\big(\bm{r}_s(\tau)\big).
\end{equation}
The ionization-time integral is evaluated by the saddle-point method,
\begin{equation}
    \frac{1}{2}\big(\bm{k}_0+\bm{A}(t_s)\big)^2=-I_p, \label{saddle_eqn}
\end{equation}
where $\bm{k}_0$ is the drift momentum, which is conserved during the Coulomb-free under-the-barrier propagation, and $\bm{k}_0+\bm{A}(\tau)$ is the corresponding kinetic momentum. The complex saddle time is $t_s=t_r+\mathrm{i}t'$. We deform the integration contour vertically from $t_s$ to $t_r$ along a line parallel to the imaginary-time axis and then propagate the trajectory in real time. The action becomes
\begin{equation}
\begin{split}
    S(t_s)&=I_p(t_s-t_i)-\frac{1}{2}\int_{t_s}^{t_r}\big(\bm{k}_0+\bm{A}(\tau)\big)^2d\tau\\
    &-\int_{t_r}^{t_f}\big(\bm{r}'(\tau)\cdot\dot{\bm{p}'}(\tau)+H(\bm{r}'(\tau),\bm{p}'(\tau),\tau)\big)d\tau.
\end{split}
\end{equation}
We evaluate the under-the-barrier action analytically and integrate the real-time action together with the equations of motion. After the pulse, each outgoing trajectory is mapped analytically to its asymptotic momentum using a Kepler hyperbola~\cite{PhysRevA.94.013415}. The tunnel-exit conditions are
\begin{equation}
    \bm{p}_{s0}(t_r)=\bm{k}_0+\bm{A}(t_r),
\end{equation}
\begin{equation}
    \bm{r}_{s0}(t_r)=\text{Re}\bigg[\int_{t_s}^{t_r}\big(\bm{k}_0+\bm{A}(\tau)\big)d\tau\bigg],
\end{equation}
where discarding the imaginary part gives a real tunnel exit and avoids the branch-cut complications associated with complex real-time trajectories~\cite{PhysRevLett.117.243003, PhysRevA.98.063423}. Applying the saddle-point approximation to the time and path integrals gives
\begin{equation}
\begin{split}
    M(\bm{p}) = -{\rm{i}}\lim_{t_f\to\infty} \sum_s\bigg[\sqrt{\frac{2\pi{\rm{i}}}{(\partial^2S(t)/\partial t^2)|_{t=t_s}}} \frac{1}{\sqrt{|J_s(t_f)|}}\\
    \times~\langle\bm{p}_0(t_s)|H_I(t_s)|\psi_0\rangle e^{{\rm{i}}S(t_s)-{\rm{i}}\frac{\pi}{2}\nu_s}\bigg].
\end{split}
\end{equation}
Here, $J_s(t)=\det[\partial\bm{p}_s(t)/\partial\bm{p}_0]$ is the stability determinant and $\nu_s$ is the Maslov index. The Maslov index accumulates when a trajectory crosses a focal point at which $J_s(t)=0$~\cite{Carlsen_2024, PhysRevLett.124.153202, LEVIT1978223, LEVIT1977165}. The equation above is schematic because the bound-state dipole has a pole at the ionization saddle. In the numerical calculation, the Gaussian saddle factor and singular dipole are treated jointly with the modified saddle-point method described in Appendix~\ref{appendix a}. The standard strong-field approximation (SFA) is recovered when $V_0(r)$ is also neglected during continuum propagation.

Evaluating the CQSFA amplitude requires the inverse mapping from a specified final momentum $\bm{p}$ to all distinct initial momenta $\bm{p}_0$. We obtain these solutions by Monte Carlo sampling followed by clustering~\cite{PhysRevLett.124.153202}, while retaining the amplitude-level stability factor $1/\sqrt{|J|}$. For a linearly polarized field and a central potential, the classical equations are cylindrically symmetric about the $x$ axis. Individual $p_\mu$ channel amplitudes need not be axisymmetric in a fixed $z$-quantized basis, but the equally weighted, $m$-summed PMD is. A trajectory contributing to the $p_x$--$p_y$ slice can therefore be chosen in that plane. We nevertheless evaluate the full three-dimensional stability determinant, including out-of-plane variations, to retain the correct semiclassical weight and Maslov phase.

\begin{figure*}[htbp]
    \centering
    \includegraphics[width=1.0\linewidth]{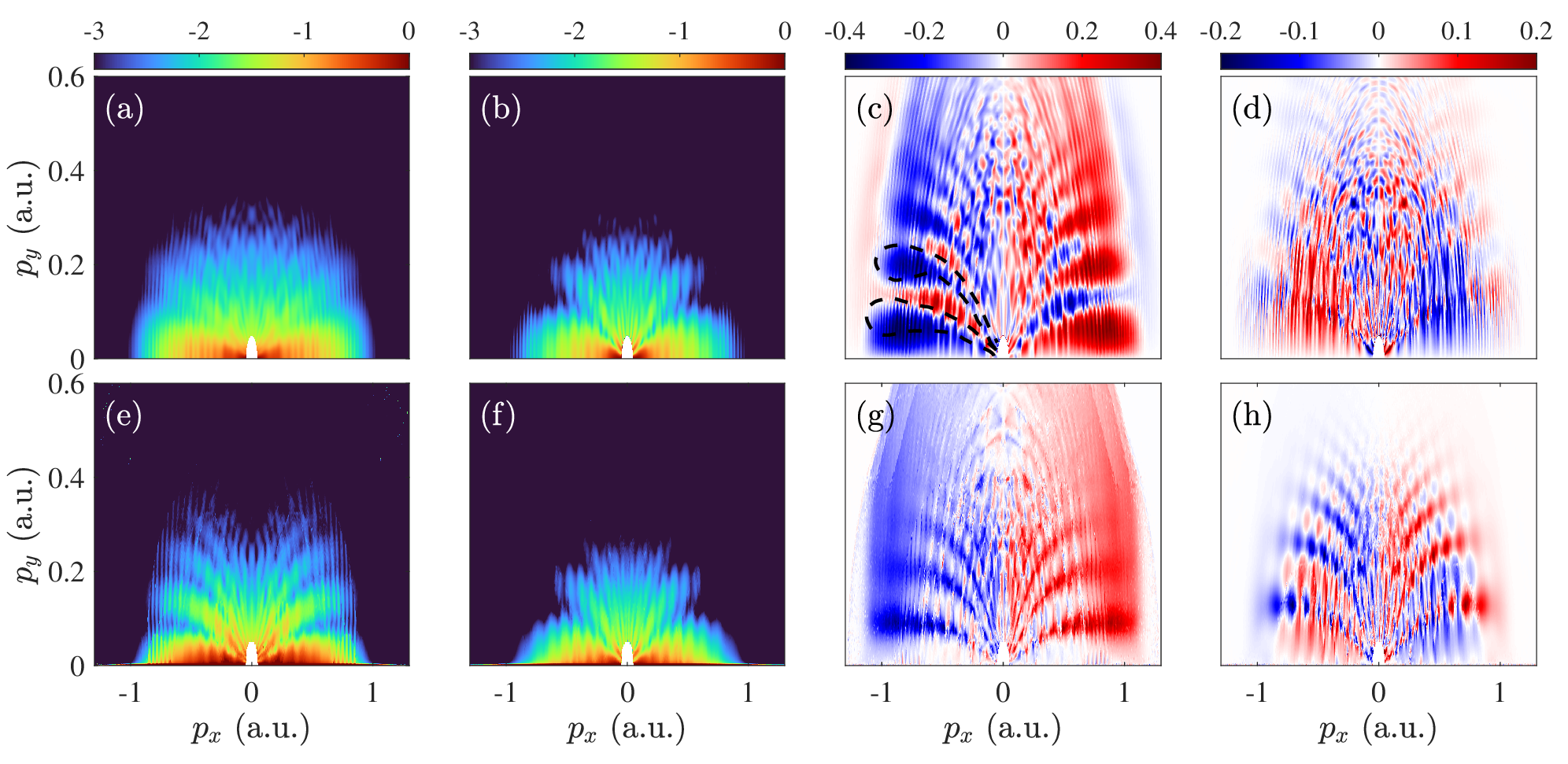}
    \caption{CEP-averaged PMDs and maps of the PST component $\zeta_z$ for $\mathrm{He^+}$ and Xe. Panels (a)--(d) show TDSE results, and panels (e)--(h) show CQSFA results. Within each row, the columns show, from left to right, the $\mathrm{He^+}$ PMD, Xe PMD, $\mathrm{He^+}$ PST, and Xe PST. The laser parameters are given in the main text, and the CEP average is taken over $0\leq\delta<2\pi$. In panel (c), the positions of the first two spider legs extracted from the CQSFA result in panel (g) are superimposed to highlight the agreement between TDSE and CQSFA.}
    \label{spin_texture}
\end{figure*}

\section{Results and discussion \label{seciv}}
The PST forms a vortex around the laser polarization axis~\cite{PhysRevLett.134.163201}. Axial symmetry relates its transverse Cartesian components by rotation, so we present $\zeta_z$ in the $p_x$--$p_y$ plane.
Figure~\ref{spin_texture} compares CEP-averaged PMDs and $\zeta_z$ maps for $\mathrm{He^+}$ and Xe. Panels (a)--(d) show TDSE results, and panels (e)--(h) show CQSFA results. The PMDs contain incoherent contributions from the unresolved $p_0$ and $p_{\pm}$ orbital channels. Both targets exhibit near-horizontal spider fringes extending along the polarization axis, and the CQSFA reproduces their principal geometry. The CQSFA PMDs also contain a glory-type axial caustic generated by Coulomb focusing. In the bare semiclassical expression, this caustic occurs when $|\partial\bm{p}_{\perp}/\partial\bm{p}_{\perp0}|=0$~\cite{PhysRevLett.121.143201, FORD1959259}; its divergence is an artifact of the nonuniform approximation.

The PSTs follow the spider geometry, with the polarization alternating between positive and negative values along adjacent legs. The qualitative agreement between the CQSFA PSTs and the TDSE benchmarks indicates that the CQSFA captures the dominant spin-texture features. The PMD caustic does not appear as a divergence in the normalized PST. The remaining discrepancies for $\mathrm{He^+}$ may arise from the neglected under-the-barrier ionic potential and from bound-state resonances, neither of which is described by the CQSFA~\cite{PhysRevA.92.043407}.

\begin{figure}[htbp]
    \centering
    \includegraphics[width=1.0\linewidth]{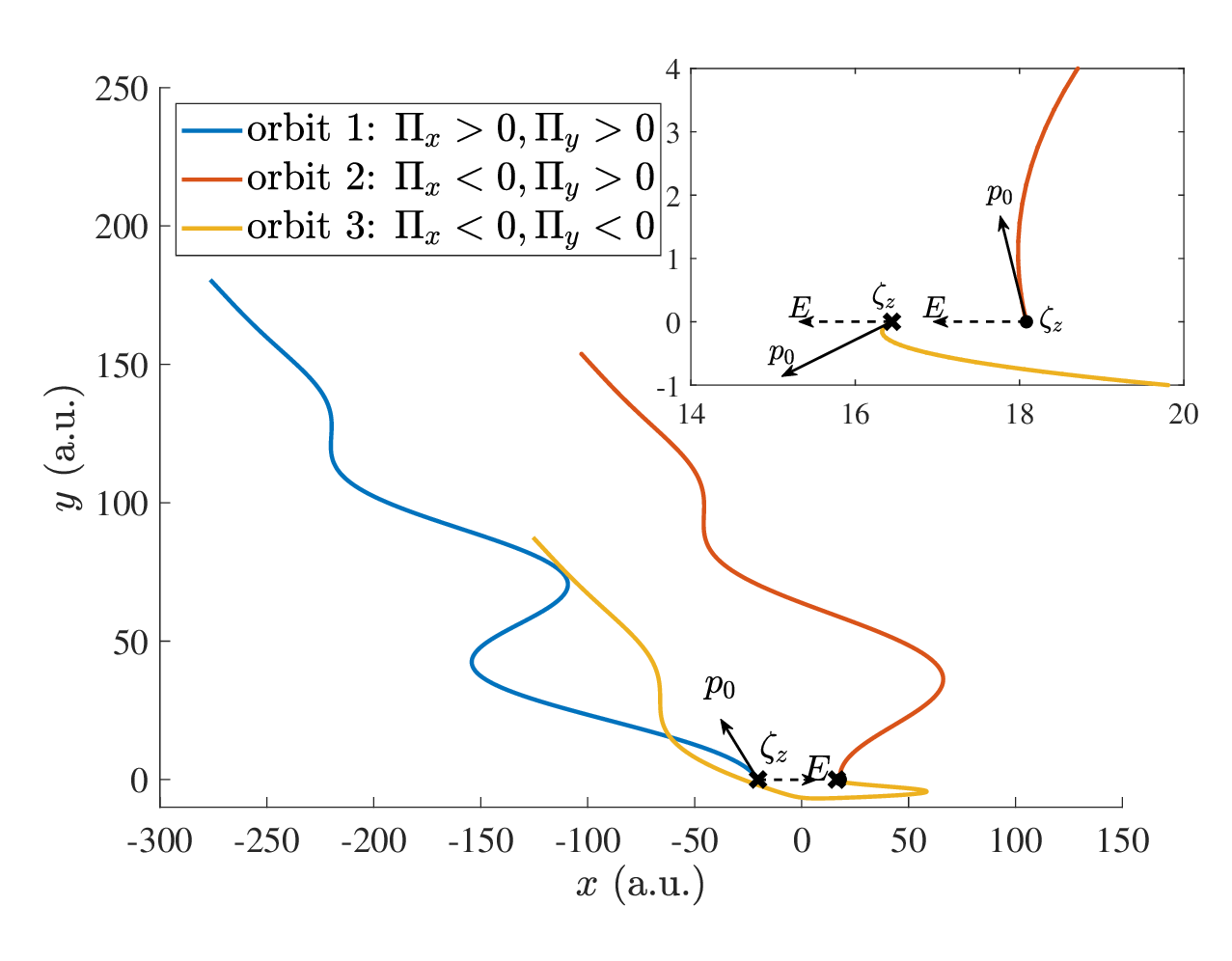}
    \caption{Illustration of the three dominant orbit classes for Xe at a final momentum $\bm{p}=(-0.3,0.25)$ a.u. Although only these three representative trajectories are shown, orbit class 4 is retained in the full calculation. At the tunnel exit, the directions of the initial momentum and electric field are indicated. The symbols $\times$ and $\bullet$ denote negative and positive values of $\zeta_z$ at the tunnel exit, respectively. The inset shows a magnified view of representative trajectories from orbit classes 2 and 3 near the tunnel exit.}
    \label{orbitclassification}
\end{figure}

\begin{table}[htbp]
\centering
\caption{Classification of orbits.}
\setlength{\tabcolsep}{6pt}
\label{table}
\begin{tabular}{cccc}
\hline\hline
Orbit class & $\Pi_x$ & $\Pi_y$ & Behavior     \\
\hline
1 & $+$ & $+$ & Direct propagation       \\
2 & $-$ & $+$ & Laser-driven deflection  \\
3 & $-$ & $-$ & Forward scattering       \\
4 & $+$ & $-$ & Backward scattering      \\
\hline\hline
\end{tabular}
\end{table}

\begin{figure}[htbp]
    \centering
    \includegraphics[width=1.0\linewidth]{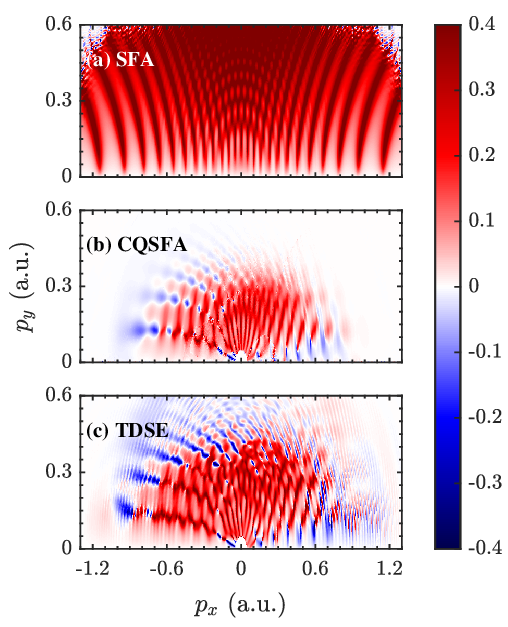}
    \caption{PST component $\zeta_z$ for Xe calculated using
    (a) SFA, (b) CQSFA, and (c) TDSE. The laser parameters are the same as those in Fig.~\ref{spin_texture}, with $\delta=0$. The CQSFA reproduces the dominant sign pattern and several principal TDSE fringe branches, whereas the SFA lacks the spin-spider modulation.}
    \label{cep=0}
\end{figure}

The $\mathrm{He^+}$ PMD contains approximately one more visible spider leg than the Xe PMD, but their overall structures remain qualitatively similar. The PSTs provide sharper contrast between the targets: the polarization of the first spider leg has opposite signs for $\mathrm{He^+}$ and Xe [Figs.~\ref{spin_texture}(c) and~\ref{spin_texture}(d)], and the Xe legs are visibly broader. In conventional SFPH, the PMD spider is generated by interference between laser-deflected and forward-scattered trajectories. Because these trajectories are released within the same quarter cycle, the resulting PMD fringes have limited sensitivity to the details of the target potential.

To identify the origin of this contrast, we use the orbit-classification scheme of Ref.~\cite{PhysRevLett.105.253002}. Four orbit classes are defined by the signs of $\Pi_x=x_0p_{fx}$ and $\Pi_y=p_{0y}p_{fy}$, where $x_0$ is the tunnel-exit coordinate along the polarization axis. The quantities $p_{0y}$, $p_{fx}$, and $p_{fy}$ are the initial transverse, final longitudinal, and final transverse momenta, respectively. Classes 1 and 2 contain short and long direct trajectories. Classes 3 and 4 contain forward- and backward-scattered trajectories, respectively. Table~\ref{table} summarizes this classification. In the momentum range considered here, $p<2\sqrt{U_p}$, class 4 is negligible and is not discussed further. Figure~\ref{orbitclassification} shows representative trajectories from classes 1--3 ending at $(p_x,p_y)=(-0.3,0.25)$ a.u.

\begin{figure*}[t]
    \centering
    \includegraphics[width=0.9\linewidth]{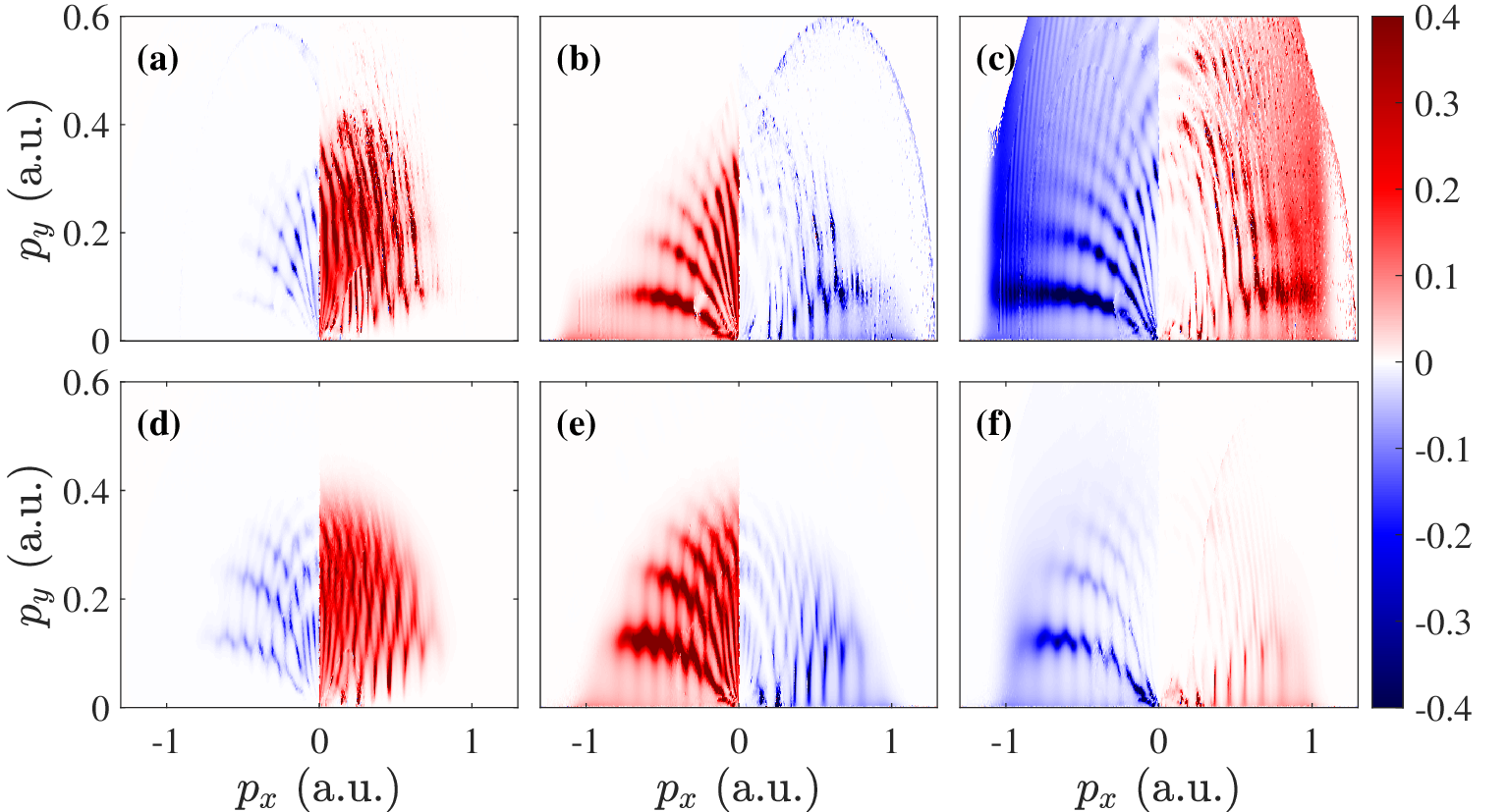}
    \caption{Diagnostic orbit-class contributions to the PST numerator at $\delta=0$, normalized by the full orbit-coherent, channel-summed PMD. Columns show orbit classes 1, 2, and 3; rows show $\mathrm{He^+}$ and Xe.}
    \label{orbit}
\end{figure*}

To resolve the orbit and channel content of the CEP-averaged maps, we next analyze the diagnostic case $\delta=0$. The one-sided features in the following maps are not intrinsic asymmetries: changing the CEP transfers the dominant contribution between momentum half-planes, whereas CEP averaging combines symmetry-related contributions. Figure~\ref{cep=0} compares $\zeta_z$ for Xe from the SFA, CQSFA, and TDSE. The CQSFA reproduces the dominant sign pattern and principal TDSE fringe branches, whereas the SFA lacks the spider modulation. In Fig.~\ref{orbit}, only the numerator of Eq.~\eqref{spin} is restricted to orbit classes 1--3; every panel retains the full orbit-coherent, channel-summed PMD in the denominator. Within the present approximation, continuum SOC and spin precession are neglected. The leading-order spin polarization assigned to each trajectory is therefore set by its tunnel-exit momentum and field direction~\cite{PhysRevLett.134.163201},
\begin{equation}
    \langle\bm{\zeta}(\bm{p}_0; j=\frac{3}{2})\rangle\approx\frac{\bm{p}_0\times \bm{E}(t_r)}{\kappa_b|\bm{E}(t_r)|}.\label{spin exit}
\end{equation}
Here, $\kappa_b=\sqrt{2I_p}$ is the bound-state momentum, distinct from the model effective charge $\kappa$ in Eq.~\eqref{psi0}. Figure~\ref{orbitclassification} illustrates the relative orientations of $\bm{p}_0$, $\bm{E}(t_r)$, and the resulting spin. Positive and negative values of $\zeta_z$ correspond to spin polarization pointing out of and into the plane, respectively. For the representative left-half-plane trajectories shown there, classes 1 and 3 have negative $\zeta_z$, whereas class 2 has positive $\zeta_z$.

At fixed CEP, orbit class 1 exhibits fan-like intracycle fringes rather than a spider [Figs.~\ref{orbit}(a) and~\ref{orbit}(d)]. Its fixed-CEP left--right asymmetry is reduced by CEP averaging. By contrast, orbit class 2 alone contains a clear spider-like PST contribution in the left half-plane. The conventional PMD spider requires interference between orbit classes 2 and 3 and cannot be produced by either class alone~\cite{PhysRevA.96.023420, Rodriguez2023}. The faint ridge at $p_x>0$ in Fig.~\ref{orbit}(b) likely reflects ambiguity in the sign-based classification: soft-recollision trajectories near a class boundary can be assigned to class 2~\cite{Rodriguez2023}.

The spider-like contribution from orbit class 2 [Figs.~\ref{orbit}(b) and~\ref{orbit}(e)] shows that interchannel coherence can generate the pattern without interorbit interference. In the plotted $p_x$--$p_y$ representation, $\zeta_z$ contains no explicit $\chi^{(0)}$ term [Eq.~\eqref{spin}]. Axial covariance relates this map, up to an overall sign fixed by the rotation convention, to $\zeta_y$ in the $p_x$--$p_z$ plane. Reflection symmetry in that plane gives $\zeta_y=\sqrt{2}\,\mathrm{Im}[\chi^{(0)*}\chi^{(+)}]/N$. The rotation convention changes only the overall sign of the extracted factor and does not alter its fringe positions.

\begin{figure*}[t]
    \centering
    \includegraphics[width=1.0\linewidth]{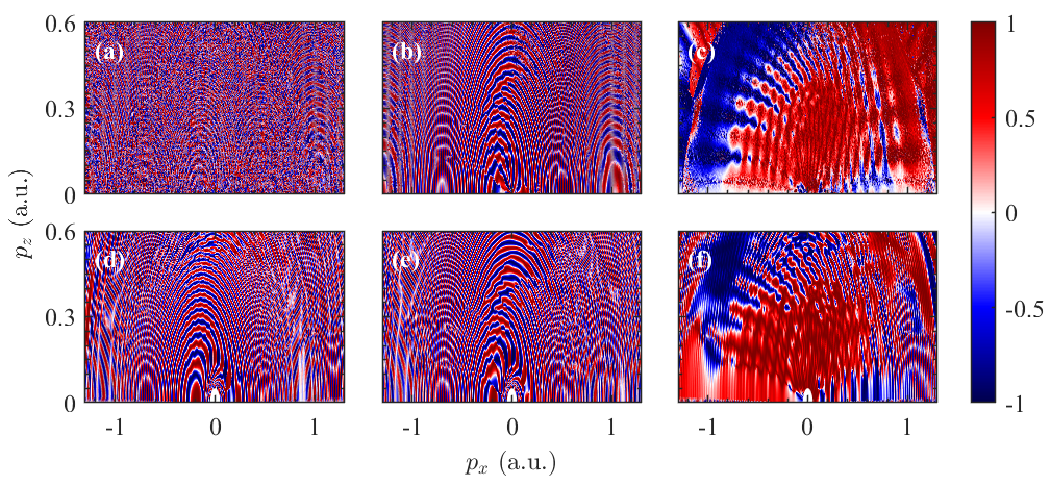}
    \caption{Channel-phase diagnostics for Xe in the $p_x$--$p_z$ plane at $\delta=0$. All quantities are constructed from total channel amplitudes coherently summed over all orbit classes. The first and second columns show $\arg\chi^{(0)}$ and $\arg\chi^{(+)}$, respectively, whereas the third column shows $\sin[\arg\chi^{(+)}-\arg\chi^{(0)}]$; the upper and lower rows correspond to the CQSFA and TDSE results, respectively. Panels (a), (b), (d), (e), and (f) are evaluated directly from the corresponding complex amplitudes. Because the low $p_0$-channel yield in the CQSFA makes the direct Monte Carlo estimate of $\arg\chi^{(0)}$ noisy, the quantity shown in panel (c) is instead extracted from $\zeta_z$ using axial covariance and the channel-amplitude normalization described in the text. No yield-based phase mask is applied. Panels (c) and (f) agree in their principal fringe positions.}
    \label{sin}
\end{figure*}

Figure~\ref{sin} analyzes the total channel amplitudes obtained by coherent summation over all orbit classes. Panels (a), (b), and (d)--(f) are evaluated directly from the corresponding complex amplitudes. The individual channel phases in panels (a), (b), (d), and (e) are convention-dependent diagnostics. The low $p_0$-channel yield in the CQSFA makes the direct Monte Carlo estimate of $\arg\chi^{(0)}$ noisy, as visible in Fig.~\ref{sin}(a). We therefore extract the quantity shown in Fig.~\ref{sin}(c) from $\zeta_z$ rather than from the direct difference of the two CQSFA channel phases. After applying axial covariance, we divide the resulting $\zeta_y$ by $\sqrt{2}|\chi^{(0)}||\chi^{(+)}|/N$ to obtain $\sin[\arg\chi^{(+)}-\arg\chi^{(0)}]$, up to an overall sign fixed by the rotation convention. No yield-based phase mask is applied, so the noisy direct $p_0$ phase remains visible in Fig.~\ref{sin}(a). The extracted CQSFA fringe positions in Fig.~\ref{sin}(c) agree with those obtained by direct TDSE evaluation in Fig.~\ref{sin}(f).

\begin{figure}[htbp]
    \centering
    \includegraphics[width=1.0\linewidth]{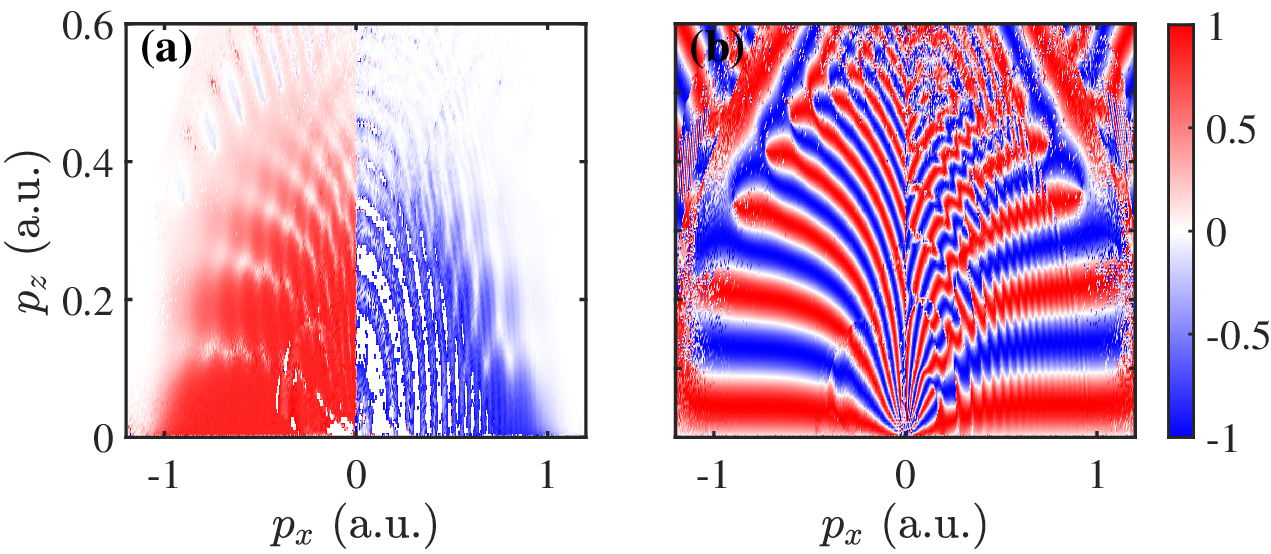}
    \caption{Phase matching for Xe at $\delta=0$. (a) Interchannel factor $\sin(\Delta S_{\rm channel})$ for the $p_+$ and $p_0$ channels within orbit class 2, extracted from the corresponding $\zeta_z$ contribution using the same symmetry and amplitude-normalization procedure as in Fig.~\ref{sin}(c). (b) Interorbit factor $\cos(\Delta S_{\rm orbit})$ evaluated directly from the class-2 and class-3 $p_+$ amplitudes. The sine and cosine arise from the imaginary PST numerator and the real PMD cross term, respectively.}
    \label{phase_matching}
\end{figure}

Because Fig.~\ref{sin} uses total amplitudes summed over all orbit classes, it supports the role of interchannel phase structure in the full PST but does not isolate a single orbit class. The single-class conclusion follows separately from the class-2 PSTs in Figs.~\ref{orbit}(b) and~\ref{orbit}(e) and from Fig.~\ref{phase_matching}(a). In the latter, the same PST-based extraction is applied to the class-2 contribution alone. Figure~\ref{phase_matching}(b), by contrast, is evaluated directly from the class-resolved $p_+$ amplitudes and describes conventional interorbit interference. The amplitude $\chi^{(\mu)}(\bm p)$ can be decomposed as
\begin{equation}
    \chi^{(\mu)}(\bm p)=\sum_{i,\{k\}_i}A_k^i d_{k,\mu}^i\exp(\mathrm{i}S_k^i),
\end{equation}
where $i=1,\ldots,4$ labels the orbit class and $\{k\}_i$ is the set of trajectories assigned to that class. The channel-independent prefactor and semiclassical action of each trajectory are represented by $A_k^i$ and $S_k^i$, respectively; $d_{k,\mu}^i$ carries the orbital-channel dependence. Within a fixed $p_+$ channel, the conventional PMD spider is governed by the total phase difference between the class-2 and class-3 sums. Within one orbit class, the spin spider is governed by the total phase difference between two orbital-channel sums. These quantities are denoted by $\Delta S_{\rm orbit}$ and $\Delta S_{\rm channel}$ in Fig.~\ref{phase_matching}; despite this notation, $\Delta S_{\rm channel}$ includes dipole and coherent-sum phases and is not a difference between the channel-independent actions $S_k^i$.

The different trigonometric factors arise from the forms of the respective observables. The class-2--class-3 cross term in the PMD is proportional to $2|M_2||M_3|\cos(\Delta S_{\rm orbit})$. The relevant PST numerator instead involves the imaginary part of an interchannel product and is proportional to $\sin(\Delta S_{\rm channel})$. The resulting alternating extrema therefore occur near $\Delta S_{\rm channel}=(n+1/2)\pi$ when the channel amplitudes overlap and vary slowly. Constructive PMD spider fringes occur near $\Delta S_{\rm orbit}=2n\pi$.

Figure~\ref{phase_matching}(a) is a phase-only diagnostic for orbit class 2 and therefore need not reproduce the normalized PST in Fig.~\ref{orbit}(e) point by point. For this class, the symmetry-mapped PST contribution can be written schematically as
\begin{equation*}
    \zeta_{\rm mapped}^{(2)}(\bm p)=
    \frac{\sqrt{2}|\chi_2^{(0)}(\bm p)||\chi_2^{(+)}(\bm p)|}{N(\bm p)}
    \sin[\Delta S_{\rm channel}^{(2)}(\bm p)].
\end{equation*}
Here, $N(\bm p)$ is the full orbit-coherent, channel-summed PMD retained in every panel of Fig.~\ref{orbit}. The interchannel phase fixes the sign-changing fringe topology, whereas the channel-amplitude envelope and momentum-dependent denominator control the observable contrast and the locations of local extrema. Consequently, the bright spider ridges in Figs.~\ref{phase_matching}(a) and~\ref{orbit}(e) need not coincide exactly. Provided that $N(\bm p)$ remains positive and finite, the normalization does not displace phase-node lines, although it can enhance, suppress, or shift neighboring extrema.

These results identify two levels of coherence. Interchannel coherence is sufficient to produce a spider-like contribution within one orbit class, whereas interorbit cross terms further modulate the full PST. The conventional PMD spider differs because it arises from interference between classes 2 and 3. In a rotated Cartesian-orbital representation, the $p_0$ channel can be viewed as a reference and the transverse $p$-orbital combination as a signal. This assignment is representation dependent; the basis-independent conclusion is that the relevant coherence is interchannel. The opposite class-2 and class-3 contributions are associated with transverse-momentum reversal during the class-3 continuum excursion, consistent with Coulomb focusing~\cite{PhysRevLett.105.253002}.

\begin{figure*}[t]
    \centering
    \includegraphics[width=0.8\linewidth]{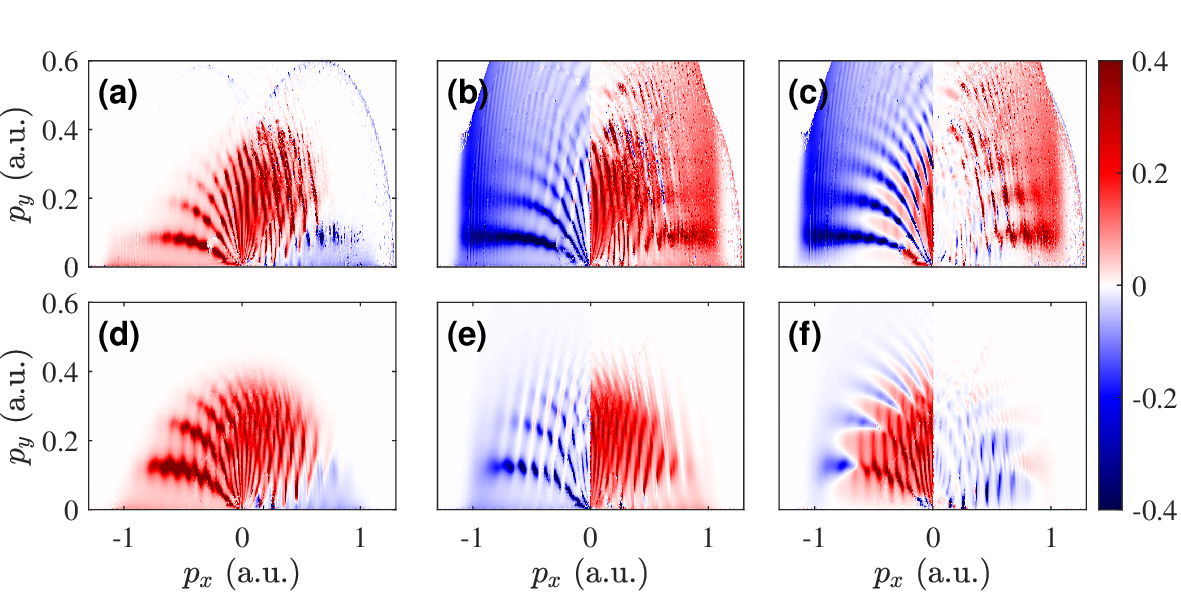}
    \caption{Coherent contributions from each orbit pair at $\delta=0$. Each panel contains both self terms and their pairwise cross term and is normalized by the full orbit-coherent, channel-summed PMD. Columns show classes 1+2, 1+3, and 2+3; rows show $\mathrm{He^+}$ and Xe.}
    \label{orbit123}
\end{figure*}

The orbit-class decomposition associates the target dependence of the first-leg sign with differences in the relative weights of the class-2 and class-3 contributions. Their coherent combination reproduces negative polarization on the left side of the $\mathrm{He^+}$ spider and positive polarization on the corresponding side of the Xe spider [Figs.~\ref{orbit123}(c) and~\ref{orbit123}(f)]. Class 3 contains trajectories whose transverse momentum reverses during continuum propagation, so this contrast is consistent with a Coulomb-focusing mechanism. Ionization of $\mathrm{He^+}$ leaves a doubly charged core with a stronger long-range Coulomb field than the core left after ionization of Xe. The resulting stronger deflection provides a plausible explanation for the difference in the balance between the class-2 and class-3 contributions. Because the two targets also differ in their bound orbitals, dipole matrix elements, and short-range potentials, the present comparison cannot isolate the effect of Coulomb focusing. It nevertheless shows that the spin spider can provide target-sensitive contrast beyond the corresponding PMD spider.

\begin{figure*}[t]
    \centering
    \includegraphics[width=1.0\linewidth]{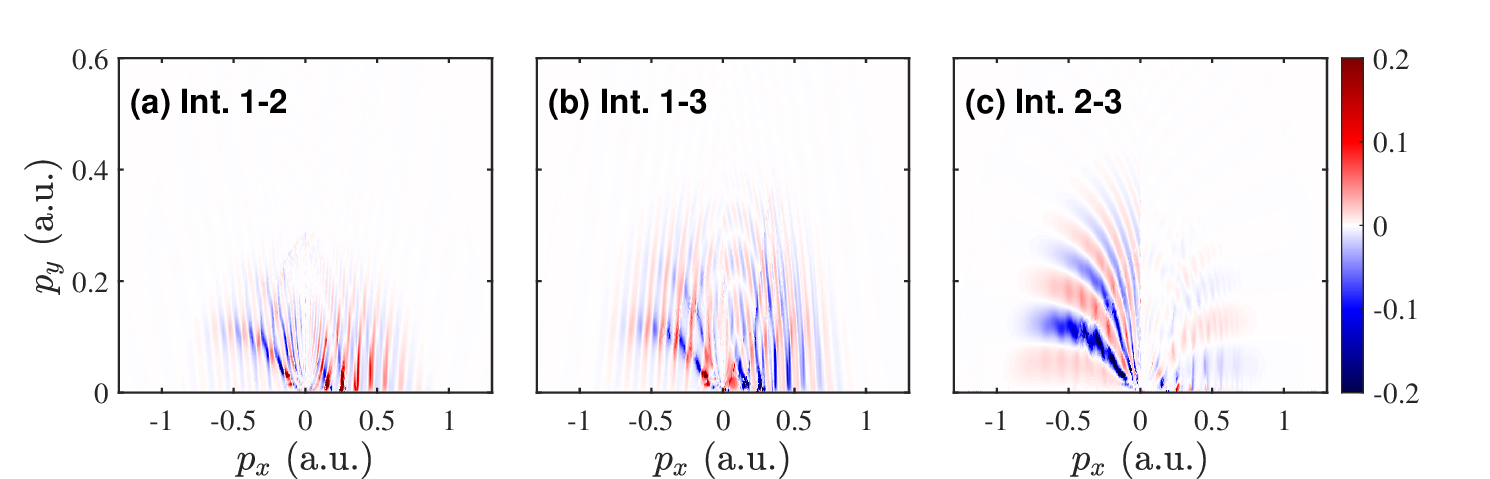}
    \caption{Diagnostic pairwise orbit-class cross terms for Xe at $\delta=0$, denoted Int.~$i$--$j$ and normalized by the full orbit-coherent, channel-summed PMD. These terms are not separately observable PSTs.}
    \label{interference}
\end{figure*}

Figure~\ref{orbit123} shows the coherent contribution from each orbit pair, including two self terms and their cross term. In Fig.~\ref{interference}, the self terms are removed to isolate the pairwise cross terms with the same orbit-coherent, channel-summed PMD normalization. This construction therefore yields a diagnostic decomposition rather than a set of separately observable PSTs. The class-1--class-2 cross term contains intracycle fringes, whereas the class-1--class-3 term contains an additional ring perpendicular to the polarization axis. The class-2--class-3 cross term forms a high-contrast spider-like pattern in both momentum half-planes. Cross terms involving class 1 are comparatively weak.

\begin{figure}[htbp]
    \centering
    \includegraphics[width=1.0\linewidth]{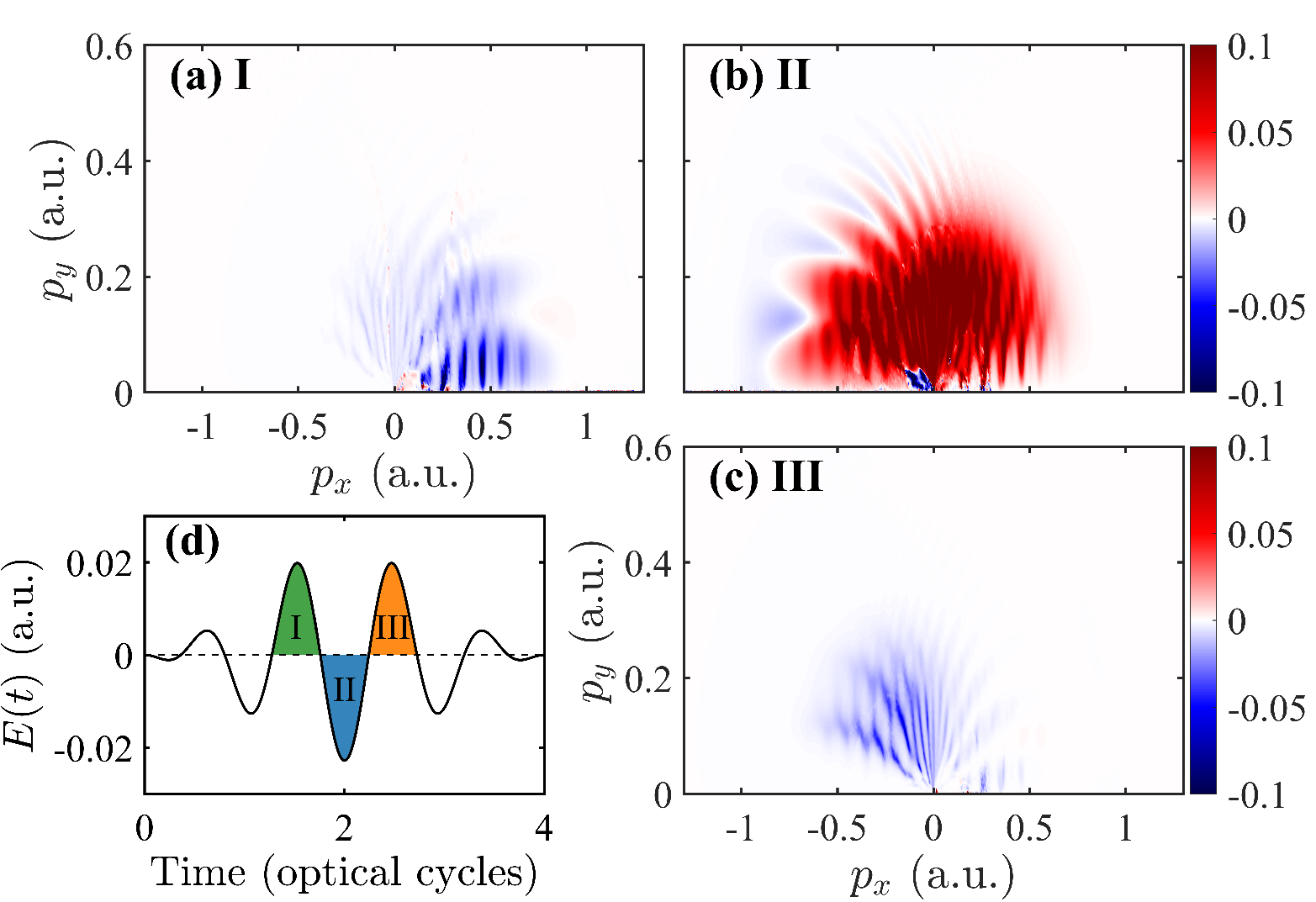}
    \caption{Panels (a)--(c) show half-cycle-resolved CQSFA contributions to the Xe PST at $\delta=0$. Each panel retains trajectories launched during the corresponding interval marked on the electric-field trace in panel (d).}
    \label{time}
\end{figure}

For $\delta=0$, the dominant single-class spider contribution appears mainly at $p_x<0$. This one-sided structure reflects the pulse envelope and the dominant ionization half-cycle, rather than an intrinsic asymmetry of the mechanism. The class-2--class-3 cross term also extends into the opposite half-plane, although with lower amplitude. Figure~\ref{time} decomposes the fixed-CEP Xe result into contributions from individual launch half-cycles. Because the tunneling rate depends exponentially on the instantaneous field, the strongest half-cycle dominates the few-cycle signal and sets the orientation of the spider. Electrons reaching opposite momentum half-planes are launched in different half-cycles and therefore experience different envelope amplitudes. The spider opens toward $p_x>0$ after a positive field maximum and toward $p_x<0$ after a negative one. Near the trailing edge of each maximum, the vector potential crosses zero and evolves opposite to $\bm{E}$. The streaking relation $\bm{p}_f\simeq-\bm{A}(t_r)$ therefore maps the dominant final momentum to the field-maximum direction~\cite{wcl3-x52t}. The half-cycle contributions have mirror-related orientations but different weights and detailed structures. The dominant spin-polarization sign of each contribution also reverses with the field direction. The cited long-pulse analysis attributes the surviving texture to recollision dynamics rather than to a simple reduction of the polarization~\cite{PhysRevLett.134.163201}. The absence of comparable spider modulation in the fixed-CEP SFA map shows that the ionic potential is required to generate the Coulomb-distorted orbit classes underlying this feature. Once that interaction is included, the labels ``direct'' and ``rescattered'' should be understood as approximate class descriptors~\cite{Maxwell2018,Rodriguez2023}.

\section{Conclusion \label{secv}}
CQSFA calculations benchmarked against TDSE results show that interchannel coherence can produce a spider-like PST contribution within a single orbit class. This mechanism differs from the conventional PMD spider, whose defining contribution is interference between orbit classes 2 and 3. Interorbit terms further modulate the full PST but are not required for the single-class pattern. The polarization along the first spider leg has opposite signs for the two targets considered. The orbit decomposition associates this contrast with Coulomb-sensitive competition between laser-deflected and forward-scattered trajectories, although a controlled potential scan would be needed to attribute the reversal uniquely to Coulomb focusing. The present conclusions apply to a single-active-electron treatment of the $j=3/2$ manifold under linearly polarized few-cycle fields, with continuum SOC and spin precession neglected.

Energy-resolved spin polarization in strong-field ionization of Xe has already been measured with time-of-flight spectroscopy and Mott polarimetry~\cite{PhysRevLett.120.043202}. Momentum-resolved spin detection with sufficient statistics could test the predicted alternating fringes, while CEP-tagged few-cycle measurements would provide subcycle sensitivity. Such measurements must account for spin-polarimetry efficiency, momentum resolution, CEP averaging, and focal-volume averaging. More broadly, spin holography provides a framework for studying the interplay among spin, orbital-channel coherence, and Coulomb-driven motion. Extending it to molecules will require explicit treatment of multicenter potentials, orientation dependence, and molecular SOC before structural or chiral imaging claims can be assessed. Combining momentum- and spin-resolved observables may ultimately provide a more complete view of ultrafast strong-field dynamics.

\begin{acknowledgments}
This work was supported by the National Natural Science Foundation of China (NSFC; Grant Nos. 12450405, 12274294, 12574378, 12574377, and 11925405). The calculations were performed on the Siyuan-1 cluster supported by the Center for High Performance Computing at Shanghai Jiao Tong University. C.-T. acknowledges support from the T.D. Lee Scholarship.
P.-L. H. acknowledges support from the Pujiang Program of the Shanghai Baiyulan Talent Plan (Grant No.~24PJA046), the Xiaomi Young Scholar Program, the Shanghai Jiao Tong University 2030 Initiative, and the Yangyang Development Fund. 
\end{acknowledgments}

\appendix

\onecolumngrid

\section{\texorpdfstring{Ionization dipole matrix elements}{Ionization dipole matrix elements} \label{appendix a}}
For a selected CQSFA model orbital $\ket{\psi_0}$, the length-gauge ionization dipole is
\begin{equation}
    d(\bm{p}) = \langle\bm{p}|\bm{r}\cdot\bm{E}|\psi_0\rangle. \label{dipole}
\end{equation}
The dipole is evaluated at the generally complex-valued saddle-point momentum. Because the applied field is polarized along $x$, only the $x$ component of $\bm r$ contributes. The plane wave is expanded in spherical harmonics as
\begin{equation}
    \langle \bm{r}|\bm{p}\rangle=\sqrt{\frac{2}{\pi}}\sum_{l=0}^{\infty}\text{i}^l j_l(pr)\sum_{m=-l}^l Y_{l,m}^*(\hat{\bm{p}})Y_{l,m}(\hat{\bm{r}}),
\end{equation}
where $j_l$ is the spherical Bessel function of the first kind. The coordinate $x$ can likewise be expanded as
\begin{equation}
    x = \sqrt{\frac{2\pi}{3}}r(Y_{1,-1}-Y_{1,1}).
\end{equation}
We use these expansions to evaluate the dipoles for the hydrogenlike model orbitals representing $\mathrm{He^+}\,2p$ and Xe $5p$ in the CQSFA.

\subsection{\texorpdfstring{$\mathrm{He^+}\,2p$}{He+ 2p} state}
For hydrogenlike $\mathrm{He^+}$ ($\kappa=2$), the exact $2p$ orbital is
\begin{equation}
    \psi_{2pm_0}(r,\theta,\phi) = \frac{1}{2\sqrt{6}} \kappa^{\frac{5}{2}}r\text{e}^{-\kappa r/2} Y_{1,m_0}(\theta, \phi).
\end{equation}
Substitution into Eq.~\eqref{dipole} yields
\begin{equation}
    d(\bm{p})= \sqrt{\frac{2}{\pi}}\sqrt{\frac{2\pi}{3}}\frac{1}{2\sqrt{6}}\kappa^{\frac{5}{2}}E_x\sum_{l=0}^{\infty}\sum_{m=-l}^{l}(-\text{i})^l Y_{l,m}(\hat{\bm{p}})\int_0^{\infty}j_l(pr)r^4\text{e}^{-\kappa r/2}dr \int Y_{l,m}^*(Y_{1,-1}-Y_{1,1})Y_{1,m_0} d\Omega. \label{2p}
\end{equation}
The identity for the integral of three spherical harmonics is
\begin{equation}
    \int Y_{l_1 m_1}^*Y_{l_2 m_2}\,Y_{l_3 m_3}d\Omega =(-1)^{m_1}\sqrt{\frac{(2l_1+1)(2l_2+1)(2l_3+1)}{4\pi}}
        \begin{pmatrix}
        l_1 & l_2 & l_3\\
        0   & 0   & 0
        \end{pmatrix}
        \begin{pmatrix}
        l_1  & l_2 & l_3\\
        - m_1 & m_2 & m_3
        \end{pmatrix}.
\end{equation}
Applying this identity, we find that the two factors with $l=1$ restrict the continuum angular momentum to $l=0$ or $2$. The sum therefore contains only $j_0$ and $j_2$.
\subsubsection{\texorpdfstring{$l=0$}{l=0}}
Substituting $j_0(pr)=\frac{\sin{pr}}{pr}$ into the radial integral, we obtain
\begin{equation}
    \int_0^{\infty}\frac{\sin(pr)}{pr}r^4\text{e}^{-\kappa r/2}dr=\frac{3\kappa(-4p^2+\kappa^2)}{(p^2+\frac{1}{4}\kappa^2)^4}.
\end{equation}
\subsubsection{\texorpdfstring{$l=2$}{l=2}}
Substituting $j_2(pr)=\frac{3\sin(pr)}{p^3r^3}-\frac{\sin(pr)}{pr}-\frac{3\cos(pr)}{p^2r^2}$ into the radial integral yields three terms,
\begin{equation}
    \begin{split}
        &\int_0^{\infty}\frac{3\sin(pr)}{p^3r^3}r^4\text{e}^{-\kappa r/2}dr-\int_0^{\infty}\frac{\sin(pr)}{pr}r^4\text{e}^{-\kappa r/2}dr-\int_0^{\infty}\frac{3\cos(pr)}{p^2r^2}r^4\text{e}^{-\kappa r/2}dr \\
        &=\frac{48\kappa}{p^2(4p^2+\kappa^2)^2}-\frac{768\kappa(-4p^2+\kappa^2)}{(4p^2+\kappa^2)^4}-\frac{48\kappa(-12p^2+\kappa^2)}{p^2(4p^2+\kappa^2)^3} \\
        &=\frac{24p^2\kappa}{(p^2+\frac{1}{4}\kappa^2)^4}.
    \end{split}
\end{equation}

These radial integrals enter the $x$-polarized dipole used in the calculations.

\subsection{Xe \texorpdfstring{$5p$}{5p} state}
We approximate the active Xe $5p$ orbital using the hydrogenlike model wavefunction
\begin{equation}
    \psi_{5pm_0}(r,\theta,\phi) = \frac{2\sqrt{\frac{2}{15}}}{46875}\kappa^{\frac{5}{2}}(-2\kappa^3r^4+90\kappa^2r^3-1125\kappa r^2+3750r)\text{e}^{-\kappa r/5}Y_{1,m_0}(\theta,\phi),
\end{equation}
where $\kappa=\sqrt{5^2\times2I_p}=4.7212$ is chosen to reproduce the Xe ionization potential $I_p=0.4458$ a.u. Substitution into Eq.~\eqref{dipole} gives
\begin{equation}
    \begin{split}
        d(\bm{p})
            &= \sqrt{\frac{2}{\pi}}\sqrt{\frac{2\pi}{3}}\frac{2\sqrt{\frac{2}{15}}}{46875}\kappa^{\frac{5}{2}}E_x\sum_{l=0}^{\infty}\sum_{m=-l}^{l}(-\text{i})^lY_{l,m}(\hat{\bm{p}}) \\
            & \times \int_0^{\infty}j_l(pr)(-2\kappa^3r^7+90\kappa^2r^6-1125\kappa r^5+3750r^4)\text{e}^{-\kappa r/5}dr \\
            & \times \int Y_{l,m}^*(Y_{1,-1}-Y_{1,1})Y_{1,m_0}d\Omega.
    \end{split}
\end{equation}
As above, the Wigner $3j$ symbols restrict the allowed values of $l$ to $0$ or $2$.

\subsubsection{\texorpdfstring{$l=0$}{l=0}}
Using the same procedure as for the $\mathrm{He^+}\,2p$ state, we obtain
\begin{equation}
    \begin{split}
        & \int_0^{\infty}\frac{\sin(pr)}{pr}(-2\kappa^3r^7+90\kappa^2r^6-1125\kappa r^5+3750r^4)\text{e}^{-\kappa r/5}dr \\
        & = \frac{-45000p^8\kappa+21600p^6\kappa^3-1814.4p^4\kappa^5+34.56p^2\kappa^7-0.1152\kappa^9}{(p^2+\frac{1}{25}\kappa^2)^7}.
    \end{split}
\end{equation}

\subsubsection{\texorpdfstring{$l=2$}{l=2}}
Substituting the explicit form of $j_2$ into the radial integral yields
\begin{equation}
    \begin{split}
        & \int_0^{\infty} \bigg[\frac{3\sin(pr)}{p^3r^3}-\frac{\sin(pr)}{pr}-\frac{3\cos(pr)}{p^2r^2}\bigg](-2\kappa^3r^7+90\kappa^2r^6-1125\kappa r^5+3750r^4)\text{e}^{-\kappa r/5}dr \\
        & = \frac{90000p^8\kappa-29520p^6\kappa^3+1814.4p^4\kappa^5-21.888p^2\kappa^7}{(p^2+\frac{1}{25}\kappa^2)^7}.
    \end{split}
\end{equation}
The radial factors have poles at $\bm{p}^2+2I_p=0$, coincident with the ionization saddle. We therefore treat the pole and stationary point jointly using the modified saddle-point method~\cite{PhysRevA.55.3760}. The substitution $\bm{p}^2=-2I_p$ is applied only to regular numerator factors; the singular denominator is retained and evaluated analytically within this treatment. Alternative forms of $\psi_0$, such as the asymptotic wavefunction~\cite{perelomov1966ionization}, mainly affect the overall magnitude of the ionization rate. The momentum dependence of the ionization rate remains largely governed by the spherical harmonics.

\twocolumngrid

\bibliography{ref}

@article{PhysRevA.88.013401,
  title = {Spin-polarized electrons produced by strong-field ionization},
  author = {Barth, Ingo and Smirnova, Olga},
  journal = {Phys. Rev. A},
  volume = {88},
  issue = {1},
  pages = {013401},
  numpages = {5},
  year = {2013},
  month = {Jul},
  publisher = {American Physical Society},
  doi = {10.1103/PhysRevA.88.013401},
  url = {https://link.aps.org/doi/10.1103/PhysRevA.88.013401}
}

@article{mao2026ultrafast,
  title={Ultrafast Ionization Dynamics Encoded in a Photoelectron Spin Torus},
  author={Mao, Xiaodan and He, Feng and He, Pei-Lun},
  journal={arXiv preprint arXiv:2604.02062},
  year={2026}
}

@article{ZHANG2023108787,
title = {QPC-TDSE: A parallel TDSE solver for atoms and small molecules in strong lasers},
journal = {Comput. Phys. Commun.},
volume = {290},
pages = {108787},
year = {2023},
issn = {0010-4655},
doi = {https://doi.org/10.1016/j.cpc.2023.108787},
url = {https://www.sciencedirect.com/science/article/pii/S0010465523001327},
author = {Zhao-Han Zhang and Yang Li and Yi-Jia Mao and Feng He}
}

@article{PhysRevA.94.013415,
  title = {Semiclassical two-step model for strong-field ionization},
  author = {Shvetsov-Shilovski, N. I. and Lein, M. and Madsen, L. B. and R\"as\"anen, E. and Lemell, C. and Burgd\"orfer, J. and Arb\'o, D. G. and T\ifmmode \mbox{\H{o}}\else \H{o}\fi{}k\'esi, K.},
  journal = {Phys. Rev. A},
  volume = {94},
  issue = {1},
  pages = {013415},
  numpages = {12},
  year = {2016},
  month = {Jul},
  publisher = {American Physical Society},
  doi = {10.1103/PhysRevA.94.013415},
  url = {https://link.aps.org/doi/10.1103/PhysRevA.94.013415}
}

@article{PhysRevLett.105.253002,
  title = {Low-Energy Structures in Strong Field Ionization Revealed by Quantum Orbits},
  author = {Yan, Tian-Min and Popruzhenko, S. V. and Vrakking, M. J. J. and Bauer, D.},
  journal = {Phys. Rev. Lett.},
  volume = {105},
  issue = {25},
  pages = {253002},
  numpages = {4},
  year = {2010},
  month = {Dec},
  publisher = {American Physical Society},
  doi = {10.1103/PhysRevLett.105.253002},
  url = {https://link.aps.org/doi/10.1103/PhysRevLett.105.253002}
}

@article{PhysRevA.92.043407,
  title = {Influence of the Coulomb potential on above-threshold ionization: A quantum-orbit analysis beyond the strong-field approximation},
  author = {Lai, X.-Y. and Poli, C. and Schomerus, H. and Faria, C. Figueira de Morisson},
  journal = {Phys. Rev. A},
  volume = {92},
  issue = {4},
  pages = {043407},
  numpages = {12},
  year = {2015},
  month = {Oct},
  publisher = {American Physical Society},
  doi = {10.1103/PhysRevA.92.043407},
  url = {https://link.aps.org/doi/10.1103/PhysRevA.92.043407}
}

@article{Carlsen_2024,
doi = {10.1088/1367-2630/ad2410},
url = {https://doi.org/10.1088/1367-2630/ad2410},
year = {2024},
month = {feb},
publisher = {IOP Publishing},
volume = {26},
number = {2},
pages = {023025},
author = {Carlsen, Mads Brøndum and Hansen, Emil and Madsen, Lars Bojer and Maxwell, Andrew Stephen},
title = {Advanced momentum sampling and Maslov phases for a precise semiclassical model of strong-field ionization},
journal = {New J. Phys.}
}

@article{PhysRevLett.124.153202,
  title = {Gouy's Phase Anomaly in Electron Waves Produced by Strong-Field Ionization},
  author = {Brennecke, Simon and Eicke, Nicolas and Lein, Manfred},
  journal = {Phys. Rev. Lett.},
  volume = {124},
  issue = {15},
  pages = {153202},
  numpages = {7},
  year = {2020},
  month = {Apr},
  publisher = {American Physical Society},
  doi = {10.1103/PhysRevLett.124.153202},
  url = {https://link.aps.org/doi/10.1103/PhysRevLett.124.153202}
}

@article{LEVIT1978223,
title = {Focal points and the phase of the semiclassical propagator},
journal = {Ann. Phys.},
volume = {114},
number = {1},
pages = {223-242},
year = {1978},
issn = {0003-4916},
doi = {https://doi.org/10.1016/0003-4916(78)90268-3},
url = {https://www.sciencedirect.com/science/article/pii/0003491678902683},
author = {S Levit and K Möhring and U Smilansky and T Dreyfus}
}

@article{LEVIT1977165,
title = {The Hamiltonian path integrals and the uniform semiclassical approximations for the propagator},
journal = {Ann. Phys.},
volume = {108},
number = {1},
pages = {165-197},
year = {1977},
issn = {0003-4916},
doi = {https://doi.org/10.1016/0003-4916(77)90355-4},
url = {https://www.sciencedirect.com/science/article/pii/0003491677903554},
author = {S Levit and U Smilansky}
}

@book{kleinert2006path,
  title={Path Integrals in Quantum Mechanics, Statistics, Polymer Physics, and Financial Markets},
  author={Kleinert, Hagen},
  year={2006},
  publisher={World Scientific Publishing Company}
}

@article{doi:10.1126/science.1198450,
author = {Y. Huismans and A. Rouzée and A. Gijsbertsen and J. H. Jungmann and A. S. Smolkowska and P. S. W. M. Logman and F. Lépine and C. Cauchy and S. Zamith and T. Marchenko and J. M. Bakker and G. Berden and B. Redlich and A. F. G. van der Meer and H. G. Muller and W. Vermin and K. J. Schafer and M. Spanner and M. Yu. Ivanov and O. Smirnova and D. Bauer and S. V. Popruzhenko and M. J. J. Vrakking},
title = {Time-Resolved Holography with Photoelectrons},
journal = {Science},
volume = {331},
number = {6013},
pages = {61-64},
year = {2011},
doi = {10.1126/science.1198450},
URL = {https://www.science.org/doi/abs/10.1126/science.1198450}}

@article{Pullen2015,
author={Pullen, Michael G. and Wolter, Benjamin and Le, Anh-Thu and Baudisch, Matthias and Hemmer, Micha{\"e}l and Senftleben, Arne and Schr{\"o}ter, Claus Dieter and Ullrich, Joachim and Moshammer, Robert and Lin, C. D. and Biegert, Jens},
title={Imaging an aligned polyatomic molecule with laser-induced electron diffraction},
journal={Nat. Commun.},
year={2015},
month={Jun},
day={24},
volume={6},
number={1},
pages={7262},
issn={2041-1723},
doi={10.1038/ncomms8262},
url={https://doi.org/10.1038/ncomms8262}
}

@article{Blaga2012,
author={Blaga, Cosmin I. and Xu, Junliang and DiChiara, Anthony D. and Sistrunk, Emily and Zhang, Kaikai and Agostini, Pierre and Miller, Terry A. and DiMauro, Louis F. and Lin, C. D.},
title={Imaging ultrafast molecular dynamics with laser-induced electron diffraction},
journal={Nature},
year={2012},
month={Mar},
day={01},
volume={483},
number={7388},
pages={194-197},
issn={1476-4687},
doi={10.1038/nature10820},
url={https://doi.org/10.1038/nature10820}
}

@article{PhysRevA.102.013109,
  title = {Holographic detection of parity in atomic and molecular orbitals},
  author = {Kang, HuiPeng and Maxwell, Andrew S. and Trabert, Daniel and Lai, XuanYang and Eckart, Sebastian and Kunitski, Maksim and Sch\"offler, Markus and Jahnke, Till and Bian, XueBin and D\"orner, Reinhard and Faria, Carla Figueira de Morisson},
  journal = {Phys. Rev. A},
  volume = {102},
  issue = {1},
  pages = {013109},
  numpages = {7},
  year = {2020},
  month = {Jul},
  publisher = {American Physical Society},
  doi = {10.1103/PhysRevA.102.013109},
  url = {https://link.aps.org/doi/10.1103/PhysRevA.102.013109}
}

@article{PhysRevA.84.043420,
  title = {Subcycle interference dynamics of time-resolved photoelectron holography with midinfrared laser pulses},
  author = {Bian, Xue-Bin and Huismans, Y. and Smirnova, O. and Yuan, Kai-Jun and Vrakking, M. J. J. and Bandrauk, Andr\'e D.},
  journal = {Phys. Rev. A},
  volume = {84},
  issue = {4},
  pages = {043420},
  numpages = {8},
  year = {2011},
  month = {Oct},
  publisher = {American Physical Society},
  doi = {10.1103/PhysRevA.84.043420},
  url = {https://link.aps.org/doi/10.1103/PhysRevA.84.043420}
}

@article{PhysRevLett.116.133001,
  title = {Probing Molecular Dynamics by Laser-Induced Backscattering Holography},
  author = {Haertelt, Marko and Bian, Xue-Bin and Spanner, Michael and Staudte, Andr\'e and Corkum, Paul B.},
  journal = {Phys. Rev. Lett.},
  volume = {116},
  issue = {13},
  pages = {133001},
  numpages = {6},
  year = {2016},
  month = {Apr},
  publisher = {American Physical Society},
  doi = {10.1103/PhysRevLett.116.133001},
  url = {https://link.aps.org/doi/10.1103/PhysRevLett.116.133001}
}

@article{wcl3-x52t,
  title = {Strong-Field Photoelectron Interferometry with Near-Single-Cycle Yb Lasers},
  author = {Hasan, Mahmudul and Tran, Phi-Hung and Gao, Jingsong and Hoang, Van-Hung and Tsai, Ming-Shian and Chen, Ming-Chang and Thumm, Uwe and Cocke, Charles Lewis and Lin, Chii-Dong and Le, Anh-Thu and Han, Meng},
  journal = {Phys. Rev. Lett.},
  volume = {135},
  issue = {26},
  pages = {263001},
  numpages = {7},
  year = {2025},
  month = {Dec},
  publisher = {American Physical Society},
  doi = {10.1103/wcl3-x52t},
  url = {https://link.aps.org/doi/10.1103/wcl3-x52t}
}

@article{Tong_2005,
doi = {10.1088/0953-4075/38/15/001},
url = {https://doi.org/10.1088/0953-4075/38/15/001},
year = {2005},
month = {jul},
publisher = {},
volume = {38},
number = {15},
pages = {2593},
author = {Tong, X M and Lin, C D},
title = {Empirical formula for static field ionization rates of atoms and molecules by lasers in the barrier-suppression regime},
journal = {J. Phys. B: At, Mol. Opt. Phys.}
}

@article{PhysRevLett.117.243003,
  title = {Laser-Driven Recollisions under the Coulomb Barrier},
  author = {Keil, Th. and Popruzhenko, S. V. and Bauer, D.},
  journal = {Phys. Rev. Lett.},
  volume = {117},
  issue = {24},
  pages = {243003},
  numpages = {5},
  year = {2016},
  month = {Dec},
  publisher = {American Physical Society},
  doi = {10.1103/PhysRevLett.117.243003},
  url = {https://link.aps.org/doi/10.1103/PhysRevLett.117.243003}
}

@article{PhysRevA.98.063423,
  title = {Treating branch cuts in quantum trajectory models for photoelectron holography},
  author = {Maxwell, A. S. and Popruzhenko, S. V. and Faria, C. Figueira de Morisson},
  journal = {Phys. Rev. A},
  volume = {98},
  issue = {6},
  pages = {063423},
  numpages = {15},
  year = {2018},
  month = {Dec},
  publisher = {American Physical Society},
  doi = {10.1103/PhysRevA.98.063423},
  url = {https://link.aps.org/doi/10.1103/PhysRevA.98.063423}
}

@article{PhysRevLett.134.163201,
  title = {Photoelectron Spin Texture in Tunneling Ionization Induced by a Linearly Polarized Laser Pulse},
  author = {He, Pei-Lun and Zhang, Zhao-Han and Hatsagortsyan, Karen Z. and Keitel, Christoph H.},
  journal = {Phys. Rev. Lett.},
  volume = {134},
  issue = {16},
  pages = {163201},
  numpages = {7},
  year = {2025},
  month = {Apr},
  publisher = {American Physical Society},
  doi = {10.1103/PhysRevLett.134.163201},
  url = {https://link.aps.org/doi/10.1103/PhysRevLett.134.163201}
}

@article{PhysRevLett.120.043201,
  title = {Energy- and Momentum-Resolved Photoelectron Spin Polarization in Multiphoton Ionization of Xe by Circularly Polarized Fields},
  author = {Liu, Ming-Ming and Shao, Yun and Han, Meng and Ge, Peipei and Deng, Yongkai and Wu, Chengyin and Gong, Qihuang and Liu, Yunquan},
  journal = {Phys. Rev. Lett.},
  volume = {120},
  issue = {4},
  pages = {043201},
  numpages = {6},
  year = {2018},
  month = {Jan},
  publisher = {American Physical Society},
  doi = {10.1103/PhysRevLett.120.043201},
  url = {https://link.aps.org/doi/10.1103/PhysRevLett.120.043201}
}

@article{Figueira,
doi = {10.1088/1361-6633/ab5c91},
url = {https://doi.org/10.1088/1361-6633/ab5c91},
year = {2020},
month = {jan},
publisher = {IOP Publishing},
volume = {83},
number = {3},
pages = {034401},
author = {Figueira de Morisson Faria, C and Maxwell, A S},
title = {It is all about phases: ultrafast holographic photoelectron imaging},
journal = {Rep. Prog. Phys.}
}

@article{PhysRevLett.94.053004,
  title = {Attosecond Probing of Vibrational Dynamics with High-Harmonic Generation},
  author = {Lein, Manfred},
  journal = {Phys. Rev. Lett.},
  volume = {94},
  issue = {5},
  pages = {053004},
  numpages = {4},
  year = {2005},
  month = {Feb},
  publisher = {American Physical Society},
  doi = {10.1103/PhysRevLett.94.053004},
  url = {https://link.aps.org/doi/10.1103/PhysRevLett.94.053004}
}

@article{PhysRevLett.133.023201,
  title = {Ultrafast Picometer-Resolved Molecular Structure Imaging by Laser-Induced High-Order Harmonics},
  author = {He, Lixin and Yuen, C. H. and He, Yanqing and Sun, Siqi and Goetz, E. and Le, Anh-Thu and Deng, Yu and Xu, Chengqing and Lan, Pengfei and Lu, Peixiang and Lin, C. D.},
  journal = {Phys. Rev. Lett.},
  volume = {133},
  issue = {2},
  pages = {023201},
  numpages = {7},
  year = {2024},
  month = {Jul},
  publisher = {American Physical Society},
  doi = {10.1103/PhysRevLett.133.023201},
  url = {https://link.aps.org/doi/10.1103/PhysRevLett.133.023201}
}

@article{PhysRevLett.71.1994,
  title = {Plasma perspective on strong field multiphoton ionization},
  author = {Corkum, P. B.},
  journal = {Phys. Rev. Lett.},
  volume = {71},
  issue = {13},
  pages = {1994--1997},
  numpages = {0},
  year = {1993},
  month = {Sep},
  publisher = {American Physical Society},
  doi = {10.1103/PhysRevLett.71.1994},
  url = {https://link.aps.org/doi/10.1103/PhysRevLett.71.1994}
}

@article{PhysRevA.49.2117,
  title = {Theory of high-harmonic generation by low-frequency laser fields},
  author = {Lewenstein, M. and Balcou, Ph. and Ivanov, M. Yu. and L'Huillier, Anne and Corkum, P. B.},
  journal = {Phys. Rev. A},
  volume = {49},
  issue = {3},
  pages = {2117--2132},
  numpages = {0},
  year = {1994},
  month = {Mar},
  publisher = {American Physical Society},
  doi = {10.1103/PhysRevA.49.2117},
  url = {https://link.aps.org/doi/10.1103/PhysRevA.49.2117}
}

@article{PhysRevA.51.1495,
  title = {Rings in above-threshold ionization: A quasiclassical analysis},
  author = {Lewenstein, M. and Kulander, K. C. and Schafer, K. J. and Bucksbaum, P. H.},
  journal = {Phys. Rev. A},
  volume = {51},
  issue = {2},
  pages = {1495--1507},
  numpages = {0},
  year = {1995},
  month = {Feb},
  publisher = {American Physical Society},
  doi = {10.1103/PhysRevA.51.1495},
  url = {https://link.aps.org/doi/10.1103/PhysRevA.51.1495}
}

@article{Hartung2016,
author={Hartung, Alexander and Morales, Felipe and Kunitski, Maksim and Henrichs, Kevin and Laucke, Alina and Richter, Martin and Jahnke, Till and Kalinin, Anton and Sch{\"o}ffler, Markus and Schmidt, Lothar Ph. H. and Ivanov, Misha and Smirnova, Olga and D{\"o}rner, Reinhard},
title={Electron spin polarization in strong-field ionization of xenon atoms},
journal={Nat. Photon.},
year={2016},
month={Aug},
day={01},
volume={10},
number={8},
pages={526-528},
issn={1749-4893},
doi={10.1038/nphoton.2016.109},
url={https://doi.org/10.1038/nphoton.2016.109}
}

@article{PhysRev.178.131,
  title = {Spin Orientation of Photoelectrons Ejected by Circularly Polarized Light},
  author = {Fano, U.},
  journal = {Phys. Rev.},
  volume = {178},
  issue = {1},
  pages = {131--136},
  numpages = {0},
  year = {1969},
  month = {Feb},
  publisher = {American Physical Society},
  doi = {10.1103/PhysRev.178.131},
  url = {https://link.aps.org/doi/10.1103/PhysRev.178.131}
}

@article{PhysRevLett.30.413,
  title = {Spin-Orbit Coupling and Photoelectron Polarization in Multiphoton Ionization of Atoms},
  author = {Lambropoulos, P.},
  journal = {Phys. Rev. Lett.},
  volume = {30},
  issue = {10},
  pages = {413--416},
  numpages = {0},
  year = {1973},
  month = {Mar},
  publisher = {American Physical Society},
  doi = {10.1103/PhysRevLett.30.413},
  url = {https://link.aps.org/doi/10.1103/PhysRevLett.30.413}
}

@article{PhysRevA.95.063410,
  title = {Producing spin-polarized photoelectrons by using the momentum gate in strong-field ionization experiments},
  author = {Liu, Kunlong and Renziehausen, Klaus and Barth, Ingo},
  journal = {Phys. Rev. A},
  volume = {95},
  issue = {6},
  pages = {063410},
  numpages = {7},
  year = {2017},
  month = {Jun},
  publisher = {American Physical Society},
  doi = {10.1103/PhysRevA.95.063410},
  url = {https://link.aps.org/doi/10.1103/PhysRevA.95.063410}
}

@article{G_G_Paulus_1994,
doi = {10.1088/0953-4075/27/21/003},
url = {https://doi.org/10.1088/0953-4075/27/21/003},
year = {1994},
month = {nov},
publisher = {},
volume = {27},
number = {21},
pages = {L703},
author = {G G Paulus and W Becker and W Nicklich and H Walther},
title = {Rescattering effects in above-threshold ionization: a classical model},
journal = {J. Phys. B: At. Mol. Opt. Phys.}
}

@article{PhysRevLett.109.013002,
  title = {Scaling Laws for Photoelectron Holography in the Midinfrared Wavelength Regime},
  author = {Huismans, Y. and Gijsbertsen, A. and Smolkowska, A. S. and Jungmann, J. H. and Rouz\'ee, A. and Logman, P. S. W. M. and L\'epine, F. and Cauchy, C. and Zamith, S. and Marchenko, T. and Bakker, J. M. and Berden, G. and Redlich, B. and van der Meer, A. F. G. and Ivanov, M. Yu. and Yan, T.-M. and Bauer, D. and Smirnova, O. and Vrakking, M. J. J.},
  journal = {Phys. Rev. Lett.},
  volume = {109},
  issue = {1},
  pages = {013002},
  numpages = {5},
  year = {2012},
  month = {Jul},
  publisher = {American Physical Society},
  doi = {10.1103/PhysRevLett.109.013002},
  url = {https://link.aps.org/doi/10.1103/PhysRevLett.109.013002}
}

@article{PhysRevLett.2.435,
  title = {Precession of the Polarization of Particles Moving in a Homogeneous Electromagnetic Field},
  author = {Bargmann, V. and Michel, Louis and Telegdi, V. L.},
  journal = {Phys. Rev. Lett.},
  volume = {2},
  issue = {10},
  pages = {435--436},
  numpages = {0},
  year = {1959},
  month = {May},
  publisher = {American Physical Society},
  doi = {10.1103/PhysRevLett.2.435},
  url = {https://link.aps.org/doi/10.1103/PhysRevLett.2.435}
}

@article{PhysRevA.110.033108,
  title = {Relativistic and spin-orbit dynamics at nonrelativistic intensities in strong-field ionization},
  author = {Maxwell, Andrew S. and Madsen, Lars Bojer},
  journal = {Phys. Rev. A},
  volume = {110},
  issue = {3},
  pages = {033108},
  numpages = {20},
  year = {2024},
  month = {Sep},
  publisher = {American Physical Society},
  doi = {10.1103/PhysRevA.110.033108},
  url = {https://link.aps.org/doi/10.1103/PhysRevA.110.033108}
}

@article{PhysRevLett.120.133204,
  title = {Direct Visualization of Valence Electron Motion Using Strong-Field Photoelectron Holography},
  author = {He, Mingrui and Li, Yang and Zhou, Yueming and Li, Min and Cao, Wei and Lu, Peixiang},
  journal = {Phys. Rev. Lett.},
  volume = {120},
  issue = {13},
  pages = {133204},
  numpages = {6},
  year = {2018},
  month = {Mar},
  publisher = {American Physical Society},
  doi = {10.1103/PhysRevLett.120.133204},
  url = {https://link.aps.org/doi/10.1103/PhysRevLett.120.133204}
}

@article{RevModPhys.81.163,
  title = {Attosecond physics},
  author = {Krausz, Ferenc and Ivanov, Misha},
  journal = {Rev. Mod. Phys.},
  volume = {81},
  issue = {1},
  pages = {163--234},
  numpages = {0},
  year = {2009},
  month = {Feb},
  publisher = {American Physical Society},
  doi = {10.1103/RevModPhys.81.163},
  url = {https://link.aps.org/doi/10.1103/RevModPhys.81.163}
}

@article{Salieres_2012,
doi = {10.1088/0034-4885/75/6/062401},
url = {https://doi.org/10.1088/0034-4885/75/6/062401},
year = {2012},
month = {may},
publisher = {IOP Publishing},
volume = {75},
number = {6},
pages = {062401},
author = {Salières, P and Maquet, A and Haessler, S and Caillat, J and Taïeb, R},
title = {Imaging orbitals with attosecond and Ångström resolutions: toward attochemistry?},
journal = {Rep. Prog. Phys.}
}

@article{RevModPhys.84.1011,
  title = {Theories of photoelectron correlation in laser-driven multiple atomic ionization},
  author = {Becker, Wilhelm and Liu, XiaoJun and Ho, Phay Jo and Eberly, Joseph H.},
  journal = {Rev. Mod. Phys.},
  volume = {84},
  issue = {3},
  pages = {1011--1043},
  numpages = {0},
  year = {2012},
  month = {Jul},
  publisher = {American Physical Society},
  doi = {10.1103/RevModPhys.84.1011},
  url = {https://link.aps.org/doi/10.1103/RevModPhys.84.1011}
}

@article{Li:16,
author = {Yang Li and Yueming Zhou and Mingrui He and Min Li and Peixiang Lu},
journal = {Opt. Express},
number = {21},
pages = {23697--23706},
publisher = {Optica Publishing Group},
title = {Identifying backward-rescattering photoelectron hologram with orthogonal two-color laser fields},
volume = {24},
month = {Oct},
year = {2016},
url = {https://opg.optica.org/oe/abstract.cfm?URI=oe-24-21-23697},
doi = {10.1364/OE.24.023697},
}

@article{PhysRevA.104.013109,
  title = {Dissecting subcycle interference in photoelectron holography},
  author = {Werby, Nicholas and Maxwell, Andrew S. and Forbes, Ruaridh and Bucksbaum, Philip H. and Faria, Carla Figueira de Morisson},
  journal = {Phys. Rev. A},
  volume = {104},
  issue = {1},
  pages = {013109},
  numpages = {12},
  year = {2021},
  month = {Jul},
  publisher = {American Physical Society},
  doi = {10.1103/PhysRevA.104.013109},
  url = {https://link.aps.org/doi/10.1103/PhysRevA.104.013109}
}

@article{PhysRevA.102.033111,
  title = {Spiral-like holographic structures: Unwinding interference carpets of Coulomb-distorted orbits in strong-field ionization},
  author = {Maxwell, Andrew S. and Faria, Carla Figueira de Morisson and Lai, XuanYang and Sun, RenPing and Liu, XiaoJun},
  journal = {Phys. Rev. A},
  volume = {102},
  issue = {3},
  pages = {033111},
  numpages = {10},
  year = {2020},
  month = {Sep},
  publisher = {American Physical Society},
  doi = {10.1103/PhysRevA.102.033111},
  url = {https://link.aps.org/doi/10.1103/PhysRevA.102.033111}
}

@article{doi:10.1126/science.1157980,
author = {M. Meckel  and D. Comtois  and D. Zeidler  and A. Staudte  and D. Pavičić  and H. C. Bandulet  and H. Pépin  and J. C. Kieffer  and R. Dörner  and D. M. Villeneuve  and P. B. Corkum },
title = {Laser-Induced Electron Tunneling and Diffraction},
journal = {Science},
volume = {320},
number = {5882},
pages = {1478-1482},
year = {2008},
doi = {10.1126/science.1157980},
URL = {https://www.science.org/doi/abs/10.1126/science.1157980}}

@article{PhysRevX.14.011015,
  title = {Laser-Induced Electron Diffraction in Chiral Molecules},
  author = {Rajak, Debobrata and Beauvarlet, Sandra and Kneller, Omer and Comby, Antoine and Cireasa, Raluca and Descamps, Dominique and Fabre, Baptiste and Gorfinkiel, Jimena D. and Higuet, Julien and Petit, St\'ephane and Rozen, Shaked and Ruf, Hartmut and Thir\'e, Nicolas and Blanchet, Val\'erie and Dudovich, Nirit and Pons, Bernard and Mairesse, Yann},
  journal = {Phys. Rev. X},
  volume = {14},
  issue = {1},
  pages = {011015},
  numpages = {25},
  year = {2024},
  month = {Feb},
  publisher = {American Physical Society},
  doi = {10.1103/PhysRevX.14.011015},
  url = {https://link.aps.org/doi/10.1103/PhysRevX.14.011015}
}

@article{PhysRevLett.109.073004,
  title = {Direct Visualization of Laser-Driven Electron Multiple Scattering and Tunneling Distance in Strong-Field Ionization},
  author = {Hickstein, Daniel D. and Ranitovic, Predrag and Witte, Stefan and Tong, Xiao-Min and Huismans, Ymkje and Arpin, Paul and Zhou, Xibin and Keister, K. Ellen and Hogle, Craig W. and Zhang, Bosheng and Ding, Chengyuan and Johnsson, Per and Toshima, N. and Vrakking, Marc J. J. and Murnane, Margaret M. and Kapteyn, Henry C.},
  journal = {Phys. Rev. Lett.},
  volume = {109},
  issue = {7},
  pages = {073004},
  numpages = {5},
  year = {2012},
  month = {Aug},
  publisher = {American Physical Society},
  doi = {10.1103/PhysRevLett.109.073004},
  url = {https://link.aps.org/doi/10.1103/PhysRevLett.109.073004}
}

@article{PhysRevLett.121.143201,
  title = {Quantum Interference of Glory Rescattering in Strong-Field Atomic Ionization},
  author = {Xia, Q. Z. and Tao, J. F. and Cai, J. and Fu, L. B. and Liu, J.},
  journal = {Phys. Rev. Lett.},
  volume = {121},
  issue = {14},
  pages = {143201},
  numpages = {5},
  year = {2018},
  month = {Oct},
  publisher = {American Physical Society},
  doi = {10.1103/PhysRevLett.121.143201},
  url = {https://link.aps.org/doi/10.1103/PhysRevLett.121.143201}
}

@article{FORD1959259,
title = {Semiclassical description of scattering},
journal = {Ann. Phys.},
volume = {7},
number = {3},
pages = {259-286},
year = {1959},
issn = {0003-4916},
doi = {https://doi.org/10.1016/0003-4916(59)90026-0},
url = {https://www.sciencedirect.com/science/article/pii/0003491659900260},
author = {Kenneth W Ford and John A Wheeler}
}

@article{PhysRevLett.108.263003,
  title = {Attosecond Time-Resolved Imaging of Molecular Structure by Photoelectron Holography},
  author = {Bian, Xue-Bin and Bandrauk, Andr\'e D.},
  journal = {Phys. Rev. Lett.},
  volume = {108},
  issue = {26},
  pages = {263003},
  numpages = {4},
  year = {2012},
  month = {Jun},
  publisher = {American Physical Society},
  doi = {10.1103/PhysRevLett.108.263003},
  url = {https://link.aps.org/doi/10.1103/PhysRevLett.108.263003}
}

@article{Meckel2014,
author={Meckel, M. and Staudte, A. and Patchkovskii, S. and Villeneuve, D. M. and Corkum, P. B. and D{\"o}rner, R. and Spanner, M.},
title={Signatures of the continuum electron phase in molecular strong-field photoelectron holography},
journal={Nat. Phys.},
year={2014},
month={Aug},
day={01},
volume={10},
number={8},
pages={594-600},
issn={1745-2481},
doi={10.1038/nphys3010},
url={https://doi.org/10.1038/nphys3010}
}

@article{Walt2017,
author={Walt, Samuel G. and Bhargava Ram, Niraghatam and Atala, Marcos and Shvetsov-Shilovski, Nikolay I. and von Conta, Aaron and Baykusheva, Denitsa and Lein, Manfred and W{\"o}rner, Hans Jakob},
title={Dynamics of valence-shell electrons and nuclei probed by strong-field holography and rescattering},
journal={Nat. Commun.},
year={2017},
month={Jun},
day={15},
volume={8},
number={1},
pages={15651},
issn={2041-1723},
doi={10.1038/ncomms15651},
url={https://doi.org/10.1038/ncomms15651}
}

@article{Khurelbaatar2024,
author={Khurelbaatar, Tsendsuren
and Heo, Jaewuk
and Yu, ShaoGang
and Lai, XuanYang
and Liu, XiaoJun
and Kim, Dong Eon},
title={Strong-field photoelectron holography in the subcycle limit},
journal={Light Sci. Appl.},
year={2024},
month={May},
day={08},
volume={13},
number={1},
pages={108},
issn={2047-7538},
doi={10.1038/s41377-024-01457-7},
url={https://doi.org/10.1038/s41377-024-01457-7}
}

@article{perelomov1966ionization,
  title={Ionization of atoms in an alternating electric field},
  author={Perelomov, AM and Popov, VS and Terent’Ev, MV},
  journal={Sov. Phys. JETP},
  volume={23},
  number={5},
  pages={924--934},
  year={1966},
  url={https://jetp.ras.ru/cgi-bin/dn/e_023_05_0924.pdf}
}

@article{PhysRevA.55.3760,
  title = {Multiphoton detachment of electrons from negative ions},
  author = {Gribakin, G. F. and Kuchiev, M. Yu.},
  journal = {Phys. Rev. A},
  volume = {55},
  issue = {5},
  pages = {3760--3771},
  numpages = {0},
  year = {1997},
  month = {May},
  publisher = {American Physical Society},
  doi = {10.1103/PhysRevA.55.3760},
  url = {https://link.aps.org/doi/10.1103/PhysRevA.55.3760}
}

@article{PhysRevLett.120.043202,
  title = {Spin and Angular Momentum in Strong-Field Ionization},
  author = {Trabert, D. and Hartung, A. and Eckart, S. and Trinter, F. and Kalinin, A. and Sch\"offler, M. and Schmidt, L. Ph. H. and Jahnke, T. and Kunitski, M. and D\"orner, R.},
  journal = {Phys. Rev. Lett.},
  volume = {120},
  issue = {4},
  pages = {043202},
  numpages = {4},
  year = {2018},
  month = {Jan},
  publisher = {American Physical Society},
  doi = {10.1103/PhysRevLett.120.043202},
  url = {https://link.aps.org/doi/10.1103/PhysRevLett.120.043202}
}

@article{Gruendeman2024,
  author = {Gr{\"u}ndeman, Elmer L. and Barb{\'e}, Vincent and Mart{\'i}nez de Velasco, Andr{\'e}s and Roth, Charlaine and Collombon, Mathieu and Krauth, Julian J. and Dreissen, Laura S. and Ta{\"i}eb, Richard and Eikema, Kjeld S. E.},
  title = {Laser excitation of the 1S--2S transition in singly-ionized helium},
  journal = {Commun. Phys.},
  year = {2024},
  month = {Dec},
  day = {19},
  volume = {7},
  number = {1},
  pages = {414},
  issn = {2399-3650},
  doi = {10.1038/s42005-024-01891-4},
  url = {https://doi.org/10.1038/s42005-024-01891-4}
}

@article{Rodriguez2023,
  Title		= {Forward and hybrid path-integral methods in photoelectron
		  holography: Sub-barrier corrections, initial sampling, and
		  momentum mapping},
  Author	= {Cruz Rodriguez, L. and Rook, T. and Augstein, B. B. and
		  Maxwell, A. S. and Figueira de Morisson Faria, C.},
  Journal	= {Phys. Rev. A},
  Volume	= {108},
  Issue		= {3},
  Pages		= {033114},
  Numpages	= {23},
  Year		= {2023},
  Month		= {Sep},
  Publisher	= {American Physical Society},
  Doi		= {10.1103/PhysRevA.108.033114},
  Url		= {https://link.aps.org/doi/10.1103/PhysRevA.108.033114}
}

@article{Maxwell2018,
  doi = {10.1088/1361-6455/aac164},
  url = {https://doi.org/10.1088/1361-6455/aac164},
  year = {2018},
  month = {may},
  publisher = {IOP Publishing},
  volume = {51},
  number = {12},
  pages = {124001},
  author = {Maxwell, A S and de Morisson Faria, C Figueira},
  title = {Coulomb-free and Coulomb-distorted recolliding quantum orbits in photoelectron holography},
  journal = {J. Phys. B: At. Mol. Opt. Phys.}
}

@article{PhysRevA.96.023420,
  title = {Coulomb-corrected quantum interference in above-threshold ionization: Working towards multitrajectory electron holography},
  author = {Maxwell, A. S. and Al-Jawahiry, A. and Das, T. and Faria, C. Figueira de Morisson},
  journal = {Phys. Rev. A},
  volume = {96},
  issue = {2},
  pages = {023420},
  numpages = {16},
  year = {2017},
  month = {Aug},
  publisher = {American Physical Society},
  doi = {10.1103/PhysRevA.96.023420},
  url = {https://link.aps.org/doi/10.1103/PhysRevA.96.023420}
}

@article{PhysRevA.90.043410,
  title = {Empirical formula for over-barrier strong-field ionization},
  author = {Zhang, Qingbin and Lan, Pengfei and Lu, Peixiang},
  journal = {Phys. Rev. A},
  volume = {90},
  issue = {4},
  pages = {043410},
  numpages = {7},
  year = {2014},
  month = {Oct},
  publisher = {American Physical Society},
  doi = {10.1103/PhysRevA.90.043410},
  url = {https://link.aps.org/doi/10.1103/PhysRevA.90.043410}
}

\end{document}